\documentclass[iop]{emulateapj}

\bibstyle{ApJ}
\usepackage{epsfig}
\usepackage{amsmath}
\usepackage{natbib}
\usepackage{graphicx}
\usepackage{longtable}

\newcommand\Msun  {${\rm M}_\odot$}
\newcommand\Lsun  {${\rm L}_\odot$}
\newcommand\msun{\rm M_{\odot}}

\newcommand\etal  {{et al.\ }}

\newcommand\Av {$\rm {A_V}$}
\newcommand\Mmax  {${\rm M}_{\rm{max}}$}

\begin{document}

\title{The Low-mass Stellar Population in L1641: Evidence for Environmental
Dependence of the Stellar Initial Mass Function}
\author{Wen-Hsin Hsu\altaffilmark{1}, Lee Hartmann\altaffilmark{1}, Lori Allen\altaffilmark{2},
Jes{\'u}s Hern{\'a}ndez\altaffilmark{3}, S. T. Megeath\altaffilmark{4}, Gregory Mosby\altaffilmark{5}, John J. Tobin\altaffilmark{6} and  Catherine Espaillat\altaffilmark{7}}
\altaffiltext{1}{Dept. of Astronomy, University of Michigan, 500 Church St., Ann Arbor, MI 48109}
\altaffiltext{2}{National Optical Astronomy Observatory, 950 North Cherry Ave, Tucson, AZ 85719}
\altaffiltext{3}{Centro de Investigaciones de Astronom{\'i}a, Apdo. Postal 264, M{\'e}rida 5101-A, 
Venezuela. }
\altaffiltext{4}{University of Toledo, 2801 West Bancroft Street, Toledo, OH 43606}
\altaffiltext{5}{University of Wisconsin - Madison, 475 N. Charter Street, Madison, WI 53706 }
\altaffiltext{6}{National Radio Astronomy Observatory, 520 Edgemont Road Charlottesville, VA 22903}
\altaffiltext{7}{Harvard-Smithsonian Center for Astrophysics, 60 Garden Street, MS-78, Cambridge, MA 02138, USA}
\shorttitle{The Low-mass Stellar Population in L1641}
\shortauthors{Hsu et al.}

\begin{abstract}

We present results from an optical photometric and spectroscopic survey of the young stellar population in L1641, 
the low-density star-forming region of the Orion A cloud south of the Orion Nebula Cluster (ONC).  Our goal is to 
determine whether L1641 has a large enough low-mass population to make the known lack of high-mass
stars a statistically-significant demonstration of environmental dependence of the upper
mass stellar initial mass function (IMF). Our spectroscopic sample consists of IR-excess objects selected from the {\em Spitzer/IRAC} 
survey and non-excess objects selected from optical photometry. We have spectral confirmation of 864 members, with another 98 
probable members; of the confirmed members, 406 have infrared excesses and 458 do not. 
Assuming the same ratio of stars with and without IR excesses in the highly-extincted regions, L1641 may contain as many as
$\sim 1600$ stars down to $\sim 0.1 \msun$, comparable within a factor of two to the the ONC. 
Compared to the standard models of the IMF, L1641 is deficient in O and early B stars to
a 3-4$\sigma$ significance level, assuming that we know of all the massive stars in L1641. 

With a forthcoming survey of the intermediate-mass stars, we will be in a better position to make a direct comparison with the 
neighboring, dense ONC, which should yield a stronger test of the dependence of the high-mass end of the stellar initial mass 
function upon environment.

\end{abstract}


\section{Introduction}

Do the most massive stars form preferentially in dense, massive environments?
Answering this question has proved to be frustratingly difficult.  The recent review by
Bastian \etal (2010)\nocite{Bastian10} concludes that there is no clear evidence for nonstandard 
stellar initial mass functions (IMFs) in
specific environments, though this issue ``clearly warrant(s) further study''.
One problem is that most of the studies of IMFs in relatively unevolved young regions have been 
conducted on young star clusters (Bastian \etal
2010; Weidner, Kroupa, \& Bonnell 2010, and references therein), because it is much more
difficult to study low-density, dispersed regions.  Perhaps the most extensive survey
to date on star formation in low-density enviroments is that of \citet{Luhman09}, 
which argued that samples of roughly 150 and 300 members of Taurus (depending
upon the region surveyed) showed an anomalous IMF, particularly in suggesting
a peak near spectral types K7-M0 not observed in other regions.  However, the low-mass
population of Taurus is insufficiently numerous to make a strong test of the high-mass end of the IMF. 

\begin{figure}
       \begin{center}
		\includegraphics[scale=0.55]{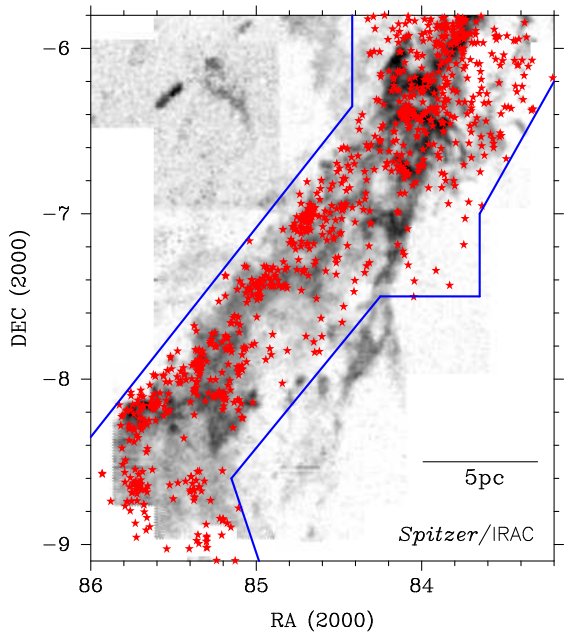}
	\end{center}
	\caption{Blue boundary lines represent the fields covered by the {\em Spitzer}/IRAC survey of Orion (Megeath et al. 2012) and the red stars show IR-excess objects identified by this survey, which are overlaid on $^{13}$CO map from \citet{Bally87}. The physical scale
	of 5 pc is also shown in the plot, assuming a distance of 420pc.  }
	\label{fig:spitzer_fields}
\end{figure}

The lower part of the Orion A molecular cloud, south of the Orion Nebula Cluster (ONC) region, appears
to be one site where a significant test of the dependence of the upper mass distribution
in a lower-density environment can be carried out.  For simplicity we call this region
``L1641'', though this is a broader use of the original Lynds cloud designation 
(as discussed further in \S~\ref{sec:early_type}, also see \citet{Allen08} for a discussion of the individual
clouds). As the lower part of the Orion A cloud is contiguous with the ONC region, it can be assumed that L1641
is at roughly the same distance as the ONC ($\sim$~420pc; \citealt{Menten07, Kim08}), allowing the 
stellar populations of the two regions to be compared directly. 
Based on previous surveys, as well as the recent {\em Spitzer/IRAC} 
survey of infrared-excess stars (see Figure~\ref{fig:spitzer_fields} for locations of the IR excess objects, data to be published in Megeath \etal 2012), 
it appears that L1641 contains a much larger pre-main sequence population than other nearby low-density regions
such as Taurus-Auriga (see, e.g \citealt{Luhman09,Rebull11}). 

The star of earliest spectral type spatially associated with the molecular gas 
south of -6.1$^\circ$~declination
is B4 \citep{Racine68}, with only an additional 12 late B stars projected on the cloud 
according to current surveys \citep{Skiff10}. The ONC, on the other hand, 
has two O stars and seven stars with spectral types earlier than B4, and its IMF 
is consistent with a Salpeter upper-mass slope \citep{Muench02, DaRio09}.
The question to be addressed is whether this apparent deficit of high-mass stars in
L1641 is the result of its being a relatively low-density environment, in contrast to
the densely clustered ONC, or whether L1641 simply does not have enough members
in comparison with the ONC (with $\sim 2000$ members; Hillenbrand \& Hartmann 1998, 
\citealt{Muench02}, \citealt{DaRio09}, \citealt{Robberto10}) that high mass stars are
probable, assuming a universal IMF.

In this paper, we present optical V and I band photometry and low resolution optical 
spectroscopy of low mass stars in L1641. We estimate that there are about 1600
 low-mass members in the region, assuming that the disk fraction is constant throughout the cloud. 
Furthermore, we show that
the average age of the stars in our sample is only slightly greater than that of the ONC region.
Thus, our findings support the idea that the comparison between L1641 and the ONC ultimately
will provide a good test of the possible environmental dependence of the upper mass IMF.
For this paper, we consider O \& B stars found in the literature and assume that there are 
no deeply embedded early B stars that we do not already know. Under this assumption, we then 
make comparisons with standard forms of the IMF which indicate that L1641 is deficient
in early B and O stars, based on the known low-mass population. In a forthcoming paper, we will
discuss our search for high-mass members and add intermediate-mass members verified by 
spectroscopy, which are currently missing from our sample.
In \S 2 we describe our observational program and the procedures used to reduce the data.
In \S 3 we present an analysis of our photometric and spectroscopic results. Then in \S 4 we 
summarize our knowledge of the low-mass population, briefly relate our results to the known
population of early-type stars, consider some uncertainties related to deciding where to
divide the region between low- and high-density areas, and make some preliminary tests
using standard models of the IMF.  Finally, we present our conclusions in \S 5.

\section{Observations and Data Reduction}\label{sec:data}

\begin{figure}
       \begin{center}
		\includegraphics[scale=0.55]{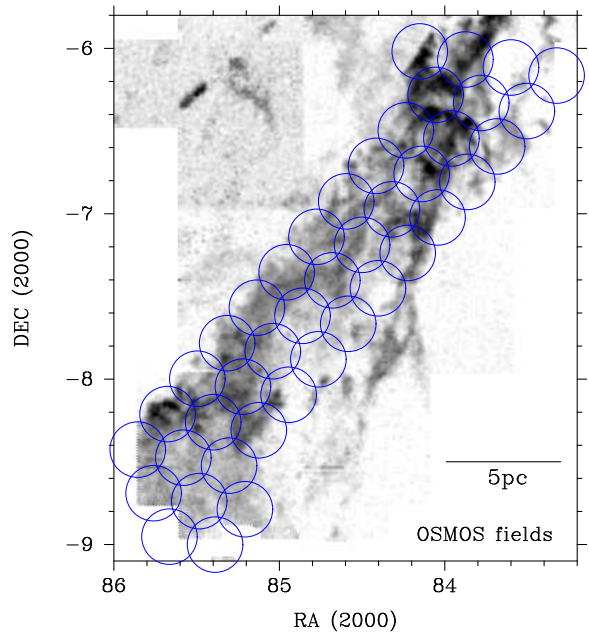}
	\end{center}
	\caption{OSMOS optical photometry fields (blue circles) overlaid on $^{13}$CO map \citep{Bally87}.The physical scale
	of 5 pc is also shown in the plot, assuming a distance of 420pc.}
	\label{fig:phot_fields}
\end{figure}

\begin{figure*}
	\begin{center}
		\includegraphics[scale=0.6]{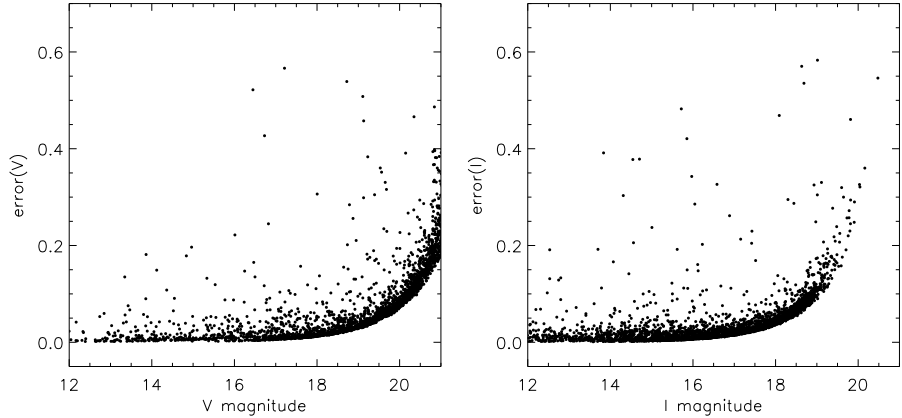}
	\end{center}
	\caption{Errors in V (left) and I (right) band photometry. The errors shown here are 
	the greater of the IRAF calculated error and the median absolute deviation.}
	\label{fig:error}
\end{figure*}

\subsection{Optical Photometry}\label{phot}
We obtained V and I band optical photometry with the Ohio State Multi-Object Spectrograph 
(OSMOS) on the MDM 2.4m Hiltner telescope \citep{Stoll10,Martini11}. 
The detector has a 
18.5\arcmin \ x 18.5\arcmin~field of view, vignetted by a 20\arcmin~diameter circle. 
We used 2 x 2 
binning, which gives a plate scale of 0.55\arcsec \ per pixel. 
The majority of L1641 from -6$^\circ$ \ to -9.2$^
\circ$ \ was covered with 41 fields (see Figure~\ref{fig:phot_fields}). The exposure sequence consists of one 
short exposure (5 seconds) and two long exposures (60 seconds) in both V and I bands. The photometry was 
obtained on the nights of Dec 7 - Dec 8 and Dec 8 - Dec 9, 2010. Most fields were observed with the exposure 
sequence once on each night, but a few fields were only observed on one night. 
The conditions were not completely photometric, with some visible cirrus at the beginning of 
each night, and the seeing ranged from 1.1\arcsec \ to 1.8\arcsec~FWHM. However, we started observing L1641 
three to four hours after twilight ended and most of the cirrus had disappeared by then. 
We observed two Landolt standard fields every two hours throughout the night for photometric calibration. 
SA92, SA95, SA98 and SA101 were used according to the time of the night.

Each CCD frame was first corrected by overscan using the IDL program proc4k written by Jason Eastman. 
Note that the original program is written for the Ohio State 4k CCD imager alone, and has been 
modified to process OSMOS data. We then performed the basic reduction following the standard procedure
using IRAF. Note that we used sky flats in the flat-field correction. 
We determined the astrometric solution with $imwcs$ in $WCSTools$, using the coordinates of stars from the 
2MASS catalog. The astrometric solutions are in general good within one pixel, or 0.55\arcsec.  
We then obtained aperture photometry with the IRAF 
$phot$ package, using an aperture of 8, or $\sim$ 3-4 times the FWHM. 
We used the Landolt fields for photometric calibrations. 
The rms departures of the standard stars from the calibration equations are $\sim$ 0.05 mag for both V and I 
bands. The MDM 4k imager used by OSMOS has a known issue of crosstalk between the four CCD 
segments. When a CCD pixel is saturated, it creates spurious point-sources in corresponding pixels on 
all other three segments. Therefore, instead of finding point sources in the images with $daofind$-like packages, 
we used the positions of all 2MASS point sources for our photometry. Non-detections and saturated stars in either bands are removed.
If a star is saturated in the long exposure but not in 
the short exposure, the short exposure is used for the photometry. Measurements 
that are affected by bleeding or crosstalk from a saturated star are manually removed. Finally, we combine all the photometry 
measurements for the same star by taking the median value.

To estimate the photometric error, we use both the error estimated by the IRAF $phot$ and the variation in 
separate observations of the same star. The IRAF $phot$ package only takes into account of Poisson 
statistics which tends to underestimate the true uncertainty. To provide a better estimate of our errors,
we take the stars that we have at least 
four long-exposure measurements to find the median absolute deviation (MAD), a robust measure of the 
deviation. Since most fields are observed on both nights and the measurements from both nights are used,
the deviation implicitly takes into account of the potential effects due to variable atmospheric extinction and 
seeing from one night to another.
If a star is saturated in the long exposures, then the short-exposure measurements (in the case 
where there are more than two short exposures) are used. We estimate the photometric error for each star 
from the greater of the $phot$ calculated error and the MAD. The results are shown in Figure~\ref{fig:error}. 
The typical errors are less than 0.05 magnitude for stars brighter than V = 19 or I = 18. This error is comparable
to the errors obtained in the calibration of Landolt stars and is most likely due to the non-photometric weather 
conditions. The typical error then increases drastically with fainter stars due to uncertainties in photon statistics. 
A small fraction of stars have errors much larger than the typical errors, which can be due to non-photometric 
conditions (cirrus in the beginning of the night) as well as intrinsic variability of the stars from one night to another. 
We think intrinsic variability contributes to the measured errors because we examined the list of stars with the 
largest errors and many of them turned out to be known variable stars. 

\begin{figure}
	\begin{center}
		\includegraphics[scale=0.5]{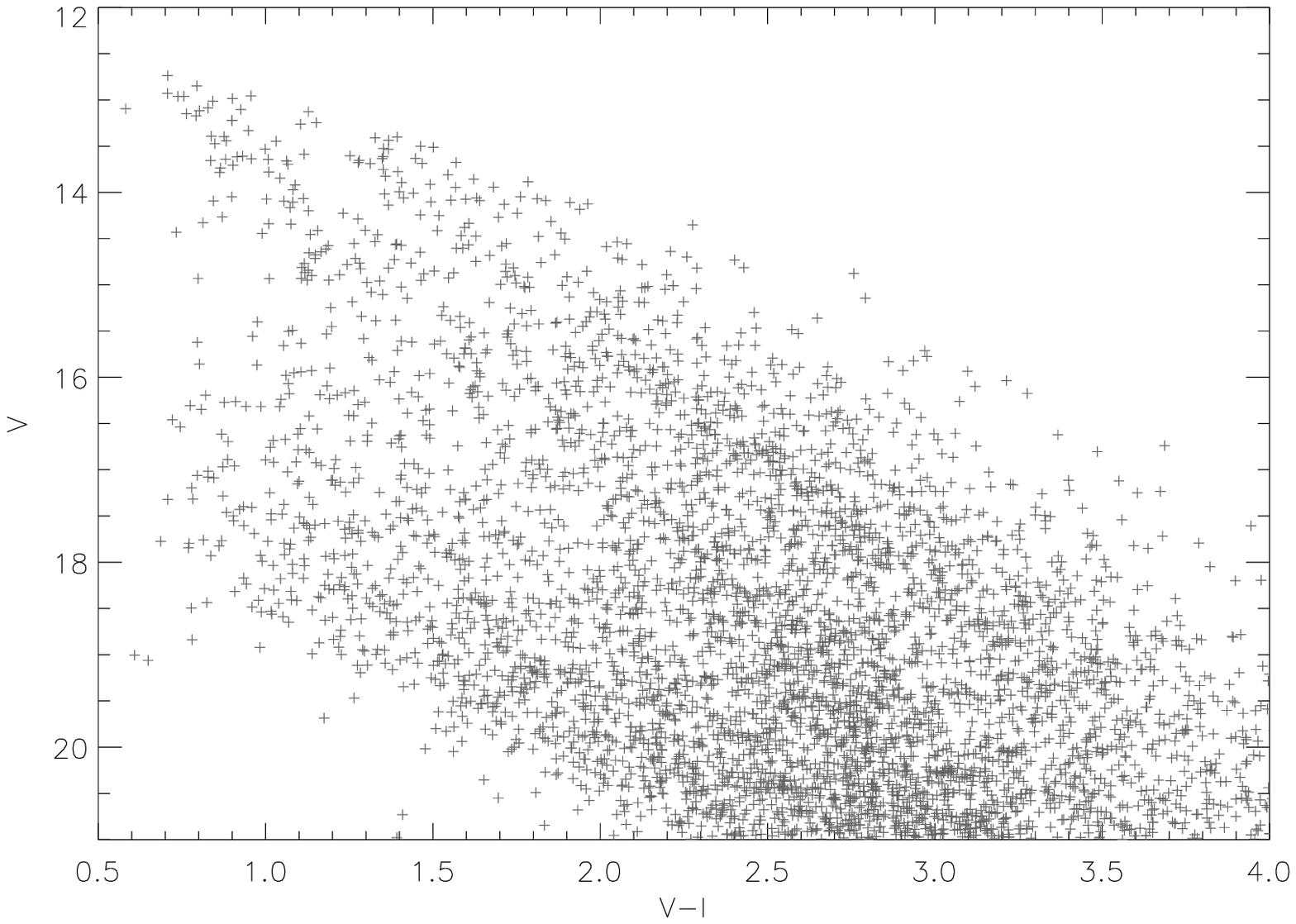}
		\includegraphics[scale=0.5]{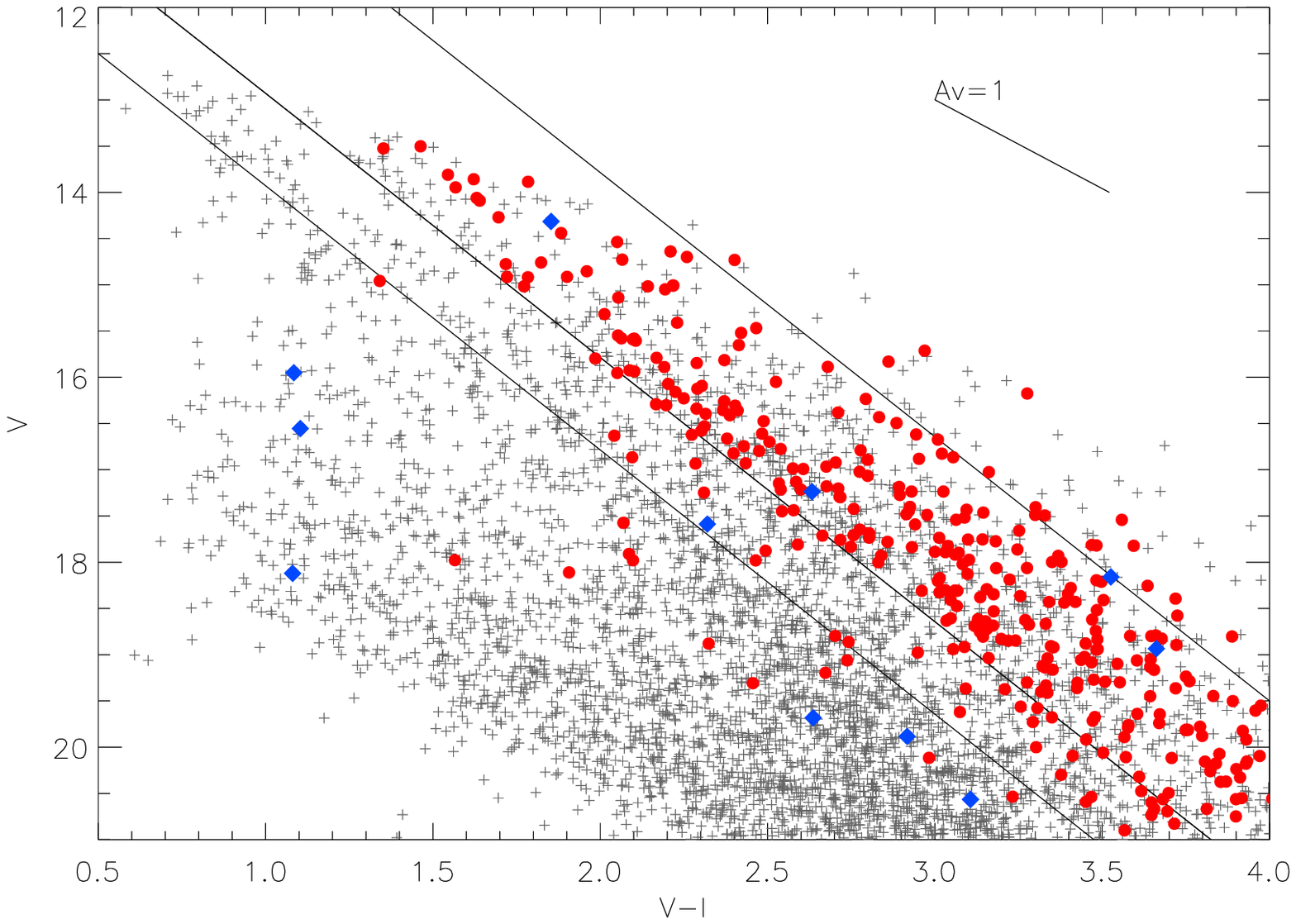}
	\end{center}
	\caption{Top: V vs. V-I color-magnitude diagram of objects in L1641, showing all the photometric data. Bottom: Same figure, with protostars (blue diamonds) and disk objects (red circles) overplotted. The YSOs with disks (red circles) lie mostly in a small region of the CMD, marking the region populated by
the pre-main sequence stars. There are 772 non-IR excess sources (grey crosses) in the same region of the CMD. This photometry was taken in Dec 2010 with OSMOS on the MDM 2.4m telescope. The extinction vector is the standard extinction taken from \citet{Cardelli89} with R$_v$ = 3.1. The parallel lines are discussed in \S~\ref{sec:lowmass}.}
	\label{fig:photometry}
\end{figure}

We were able to obtain photometry of stars with V and I magnitude between 12 and 22 mags. However, 
from Figure~\ref{fig:error}, we know that the typical error of a V = 21 mag star is about 0.2 mag and 
the error increases rapidly for fainter objects. 
We therefore limit our photometric sample to stars with 
V magnitudes between 12.5 and 21, I magnitudes between 12 and 20.5 and V-I between 0.5 and 4 mags.
with the typical error of a 21 magnitude star being 0.2 mags.

Figure~\ref{fig:photometry} shows the V vs. V-I color-magnitude diagram of the sample mentioned 
above. The top panel shows all the stars in our sample, regardless of its IR-excess and the bottom
panel shows the same plot, with the class I protostars shown in blue diamonds and the class II's
(IR excess objects with disks) shown in red circles. 

We note that most of the class II stars in Figure~\ref{fig:photometry} fall 
in a small region on the CMD, which we approximate with the top two parallel lines.
The bottom of the YSO region is traced by a line through the two points (V-I, V) = (0.5, 11.5) and (4, 21.5), 
and the top is 2 magnitudes above it in V at a given V-I. 
We expect that the Class III's (pre-main sequence objects with no IR excess) have similar ages and distances and therefore 
fall in the same region of the CMD. However, not all the non-IR excess objects in this region are 
members. We therefore target them in our optical spectroscopy surveys to confirm their membership. 

\begin{figure}
       \begin{center}
		\includegraphics[scale=0.55]{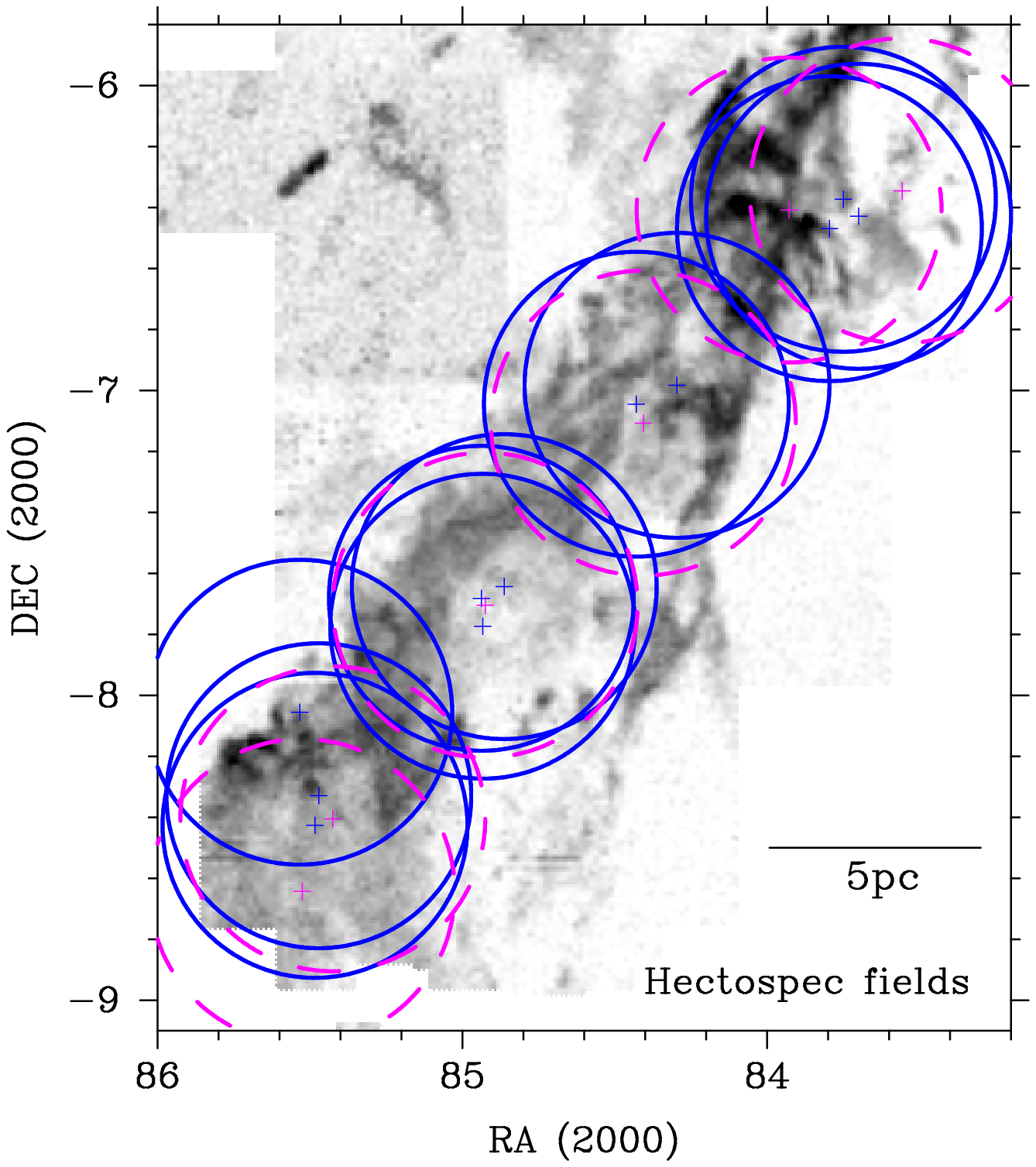}
		\includegraphics[scale=0.55]{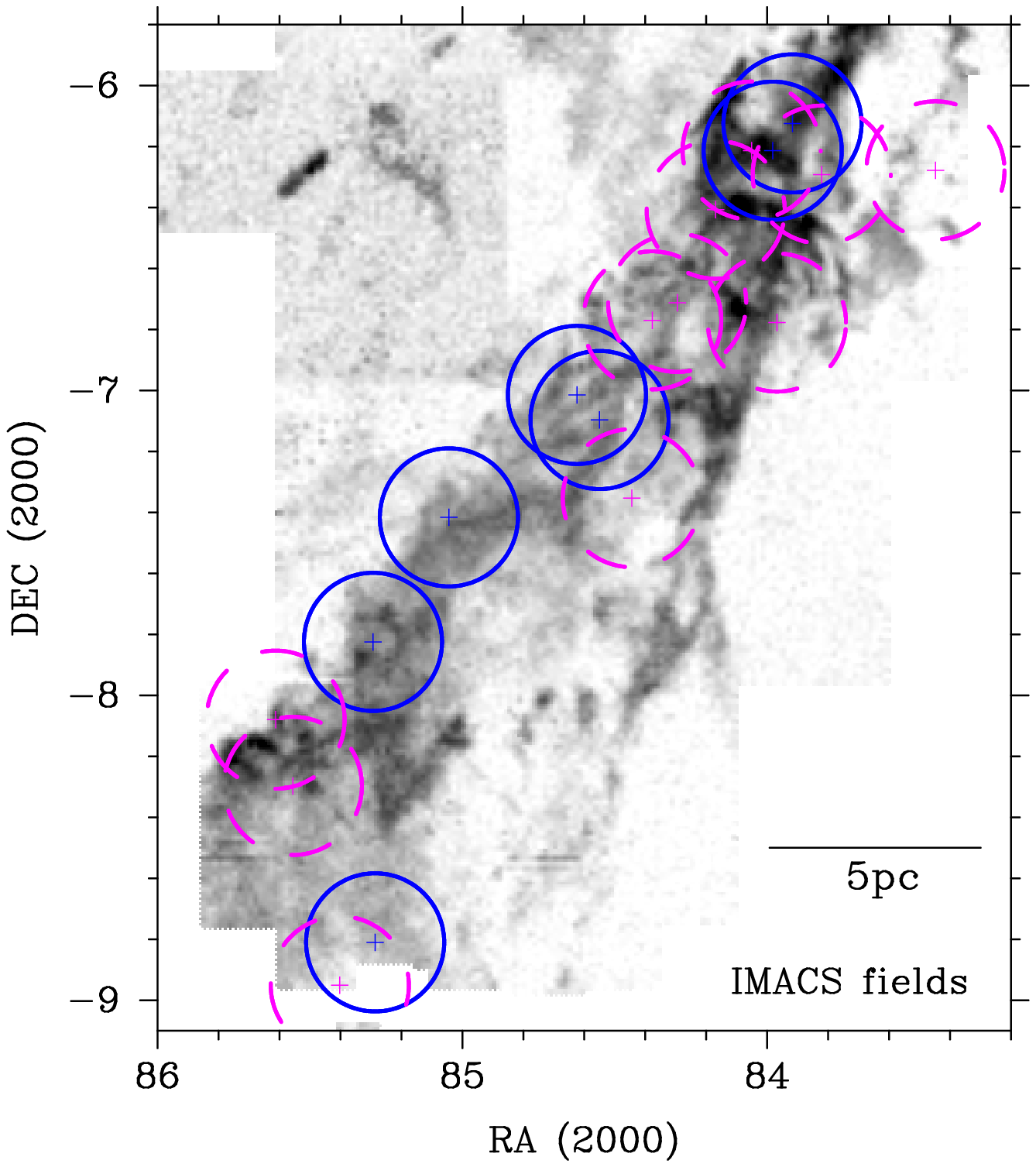}
	\end{center}
	\caption{Optical spectroscopy fields overlaid on $^{13}$CO map from \citet{Bally87}. The Hectospec fields (top) are chosen to maximize number of IR excess objects observed. The IMACS fields (bottom) sample are either IR-selected (blue solid circles) or optical-selected (magenta dashed circles). The IMACS fields are mainly used to compliment the Hectospec fields in crowded regions. The physical scale of 5 pc is also shown in the plot, assuming a distance of 420pc.}
	\label{fig:spec_fields}
\end{figure}

\subsection{Optical Spectra}\label{sec:data_spec}

There are two parts to our optical spectroscopic sample. The first part is the sample of 
infrared-excess stars from the {\em Spitzer} survey of Megeath \etal (2012). 
Figure~\ref{fig:spitzer_fields} shows fields covered by the {\em Spitzer} survey and the 
positions of the IR-excess stars. The {\em Spitzer} survey identified 166 protostars and 
723 disk objects in L1641 (defined as the field outlined in Figure~\ref{fig:spitzer_fields} 
but stops at -6$^\circ$.) Note that only 8\% of the protostars and 48\% of the disk objects 
are found in our optical photometric sample and are therefore sufficiently bright for optical spectroscopy.
The second part of our sample is non-IR excess objects selected 
based on their optical photometry. There are a total of 772 non-excess stars that satisfy
the optical photometry selection and we were able to target 95\% of these objects in our 
spectroscopy. The sample is, however, biased against heavily-reddened objects. 

The spectra were obtained with multi-aperture spectrographs covering extended fields of view. 
Approximately 75\% of our spectra were obtained with Hectospec, 
the fiber-fed multiobject spectrograph on the 6.5m MMT on Mt. Hopkins \citep{Fabricant05}. It has 300 fibers that can be placed within a 
1$^\circ$\ diameter field. Each Hectospec fiber subtends 1.5\arcsec~on the sky. The observations were 
taken with 270 line mm$^{-1}$ grating, providing $\sim$ 6.2\AA \ resolution and spectral coverage of 
3650 to 9200\AA. With 17 observations from 2006 to 2011, a total of 2369 stars were observed. 
The total on-target exposure times are 45 to 50 minutes for most fields, with the exception that the 
Jan 30, 2009 field was observed for 22.5 minutes.
The top panel of Figure~\ref{fig:spec_fields} shows fields of the Hectospec observations. Targets for observing runs
from 2006 to 2009 are selected based on their IR colors (fields shown in blue) and the targets of the 2011 run 
are selected based on their optical photometry (fields shown in magenta, see \S~\ref{phot} for selection criteria).
We first maximized the number of targets that satisfy either the IR excess or optical selection criteria, and then filled in the 
remaining fiber positions with other stars in the field. 

The Hectospec data were reduced through the standard 
Hectospec data reduction pipeline \citep{Mink07}, except for sky background subtraction. The 
pipeline assumes that the sky background does not vary significantly with position on the sky and uses 
``sky fibers", or fibers that point to empty portions of the sky, to correct for the sky background. However, 
in a star-forming region, nebulosity may result in significantly varying sky background and therefore bad 
estimates of nebular lines. Our sky subtraction takes into account the spatial variations of the sky background. 
Some of the observations were taken along with offset sky spectra, 5\arcsec \ apart from the star, in which case we subtracted
the offset sky spectra from the science target spectra. When the offset sky spectra were not taken, we used the closest 3 
to 5 sky fibers as the average sky and subtract it from the science spectra. We tested our sky subtraction methods with
sky fibers (so a perfect sky subtraction would leave a spectrum with only noises around 0) and found that  
the quality of the resulting spectra is significantly improved with the offset sky spectra, with most of the H$\alpha$ 
emission and other sky lines accounted for whereas the nearest 3 - 5 sky fibers improves the overall sky 
subtraction (especially night-sky emission lines in the near-IR) but does not 
 correct for the nebular lines very well. 
 
Another 25\% of our spectra were obtained with IMACS, the Inamori-Magellan Areal Camera \& Spectrograph 
on the Magellan Baade telescope \citep{Bigelow03}. We used IMACS f/2 camera in multi-slit spectroscopy mode with the
300 line grism at a blaze angle of 17.5$^\circ$. With a 0.6\arcsec \ slit, this 
configuration yields a resolution of 4\AA \ and spectral coverage of approximately 4000 to 9000\AA \ 
(stars close to the edge of the fields may not have full spectral coverage). From 17 observations in 2010 and 2011, 
a total of 715 stars were observed. The standard observation time for each field is 5 x 10 minutes, but we increased
the time to 6 x 10 minutes for a few observations at higher airmasses. The bottom panel of Figure~\ref
{fig:spec_fields} shows the fields of the IMACS observations. The targets of the first IMACS run were 
selected based on their IR colors (shown in blue), and the targets of the second run are selected based 
on their optical photometry (shown in magenta, see \S~\ref{phot} for selection criteria). The IMACS 
data were reduced with COSMOS, the Carnegie Observatories System for MultiObject Spectroscopy, 
following the standard cookbook. 

\begin{figure*}
       \begin{center}
		\includegraphics[scale=0.8]{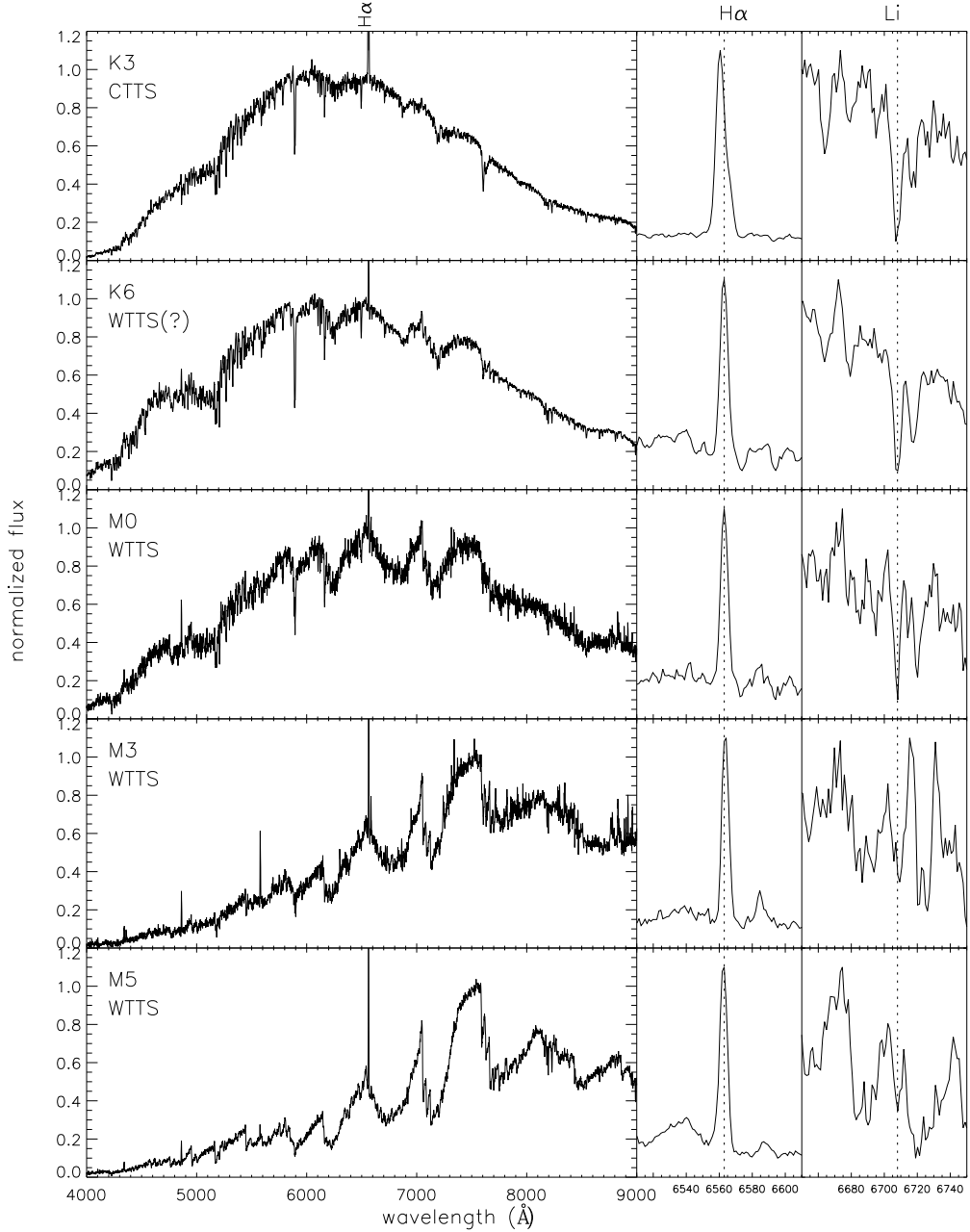}
	\end{center}
	\caption{Examples of spectra from Hectospec with a zoom-in view around H$\alpha$ ($\lambda$ 6563) and Li I ($\lambda$ 6707). From top to bottom, the spectra are spectral typed: K3, K6, M0, M3 and M5. The [O $I$] line at 5577\AA\ is a prominent sky line. The red end of the M type spectra are affected by series of sky lines. Their TTS type are classified using the definition given by \citet{WhiteBasri03}. The H$\alpha$ equivalent width in the K6 star is bordering the cutoff in the definition. }
		\label{fig:spec_Hecto}
\end{figure*}

\begin{figure*}
       \begin{center}
		\includegraphics[scale=0.8]{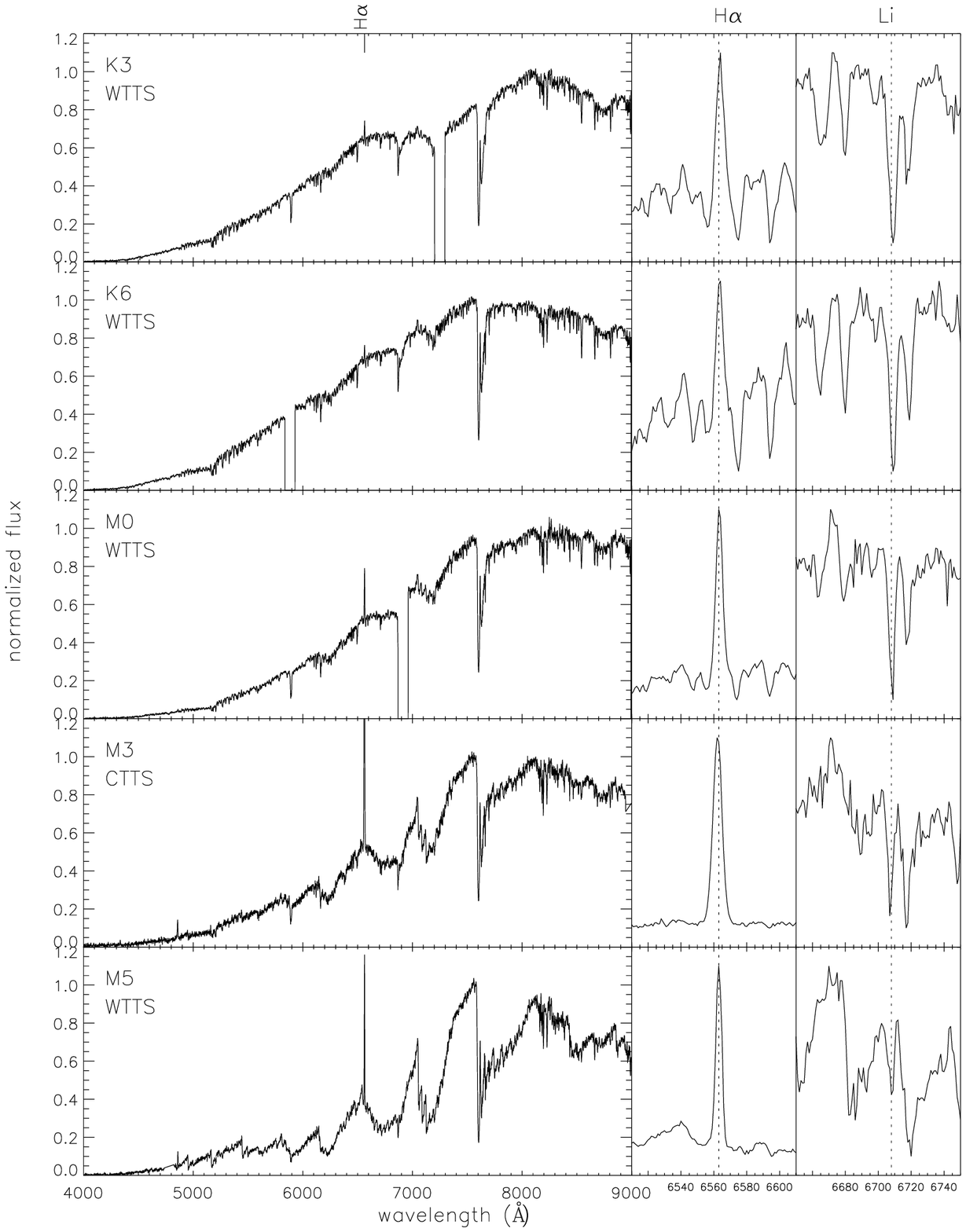}
	\end{center}
	\caption{Examples of spectra from IMACS with a zoom-in view around H$\alpha$ ($\lambda$ 6563) and Li I ($\lambda$ 6707). From top to bottom, the spectra are spectral typed: K3, K6, M0, M3 and M5. }
	\label{fig:spec_IMACS}
\end{figure*}

Under good observing conditions, we were able to spectral-type stars as faint as V $\sim$ 21. However, the faintest magnitude 
we can spectral type depends on the seeing and the instrument used.

\section{Results}
\subsection{Photometry}\label{result_phot}
Figure~\ref{fig:photometry} shows the $V$ vs. $V-I$ color-magnitude diagram of the stars between 
$V-I \sim 0.5 - 4$ and $V < 21$. There are a total of 4475 stars in this photometric sample. 
Correlating the optical photometry and the IR-excess classification \citep{Gutermuth09}, we mark the 
IR-excess objects in our photometry (bottom panel). 13 of the stars were protostars (blue diamonds) 
and 347 of them were YSOs with disks (red circles). Compared to the number of IR-excess objects 
identified in the $Spitzer$ survey, only 8\% of the protostars and 48\% of the disk objects are found
in our photometric sample. The optical photometry fields cover slightly less area than 
the $Spitzer$ survey, but cover the 95\% of the IR-excess objects. It is not surprising that we do not find most of the protostars as they are 
highly embedded in their envelopes. We were only able to identify half of the stars with disk excesses in the optical photometry 
due to extinction (see \S~\ref{extinction}). 

\subsection{Spectra and Spectral Types}\label{result_spec}
\begin{figure*}
	\begin{center}
		\includegraphics[scale=0.8]{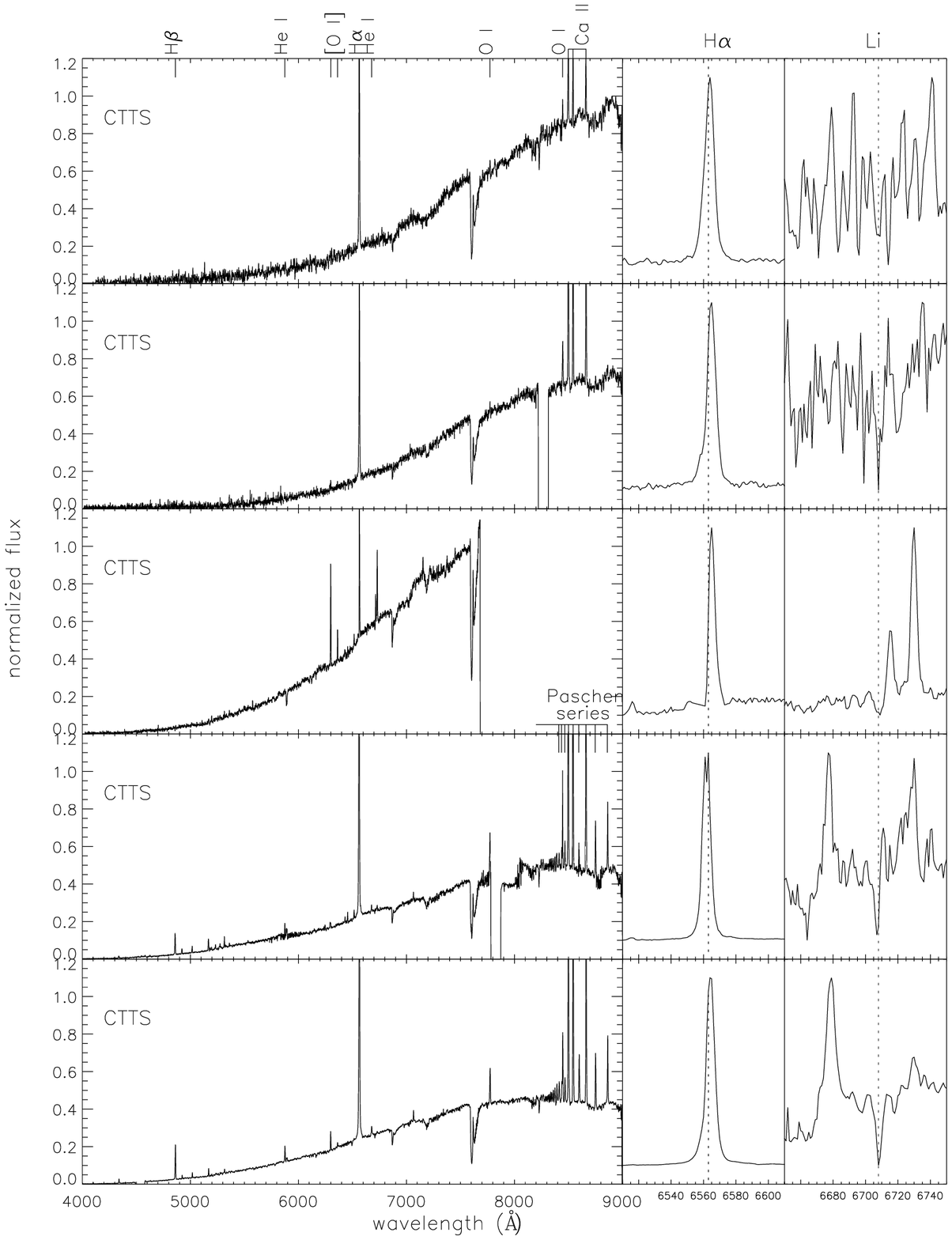}
	\end{center}
	\caption{Example IMACS spectra of rapid accretors with a zoom-in view around H$\alpha$ ($\lambda$ 6563) and Li I ($\lambda$ 6707). The spectra are highly-veiled and therefore makes spectral typing difficult. The main emission lines have been marked.}
	\label{fig:spec_acc}
\end{figure*}

Figure~\ref{fig:spec_Hecto} and~\ref{fig:spec_IMACS} show some example spectra of confirmed 
members from Hectospec and IMACS, respectively, arranged by their spectral types. The 
H$\alpha$ \ equivalent width criteria by \citet{WhiteBasri03} is used to determine whether they are 
classical T Tauri stars (CTTS) or weakline T Tauri stars (WTTS).
Note that IMACS has very little transmission in the blue whereas the transmission of Hectospec 
is more uniform across the spectral 
range. The [O I] line at 5577\AA \ is very prominent in some Hectospec data but is actually a 
sky line that was not correctly subtracted. In this case, the H$\alpha$ emission is also likely 
affected by bad sky subtraction.  Figure~\ref{fig:spec_acc} show IMACS spectra of rapidly-
accreting stars. They not only have a high H$\alpha$ equivalent width, but also have other emission 
lines such as [O I], He I and the infrared Ca triplet. The last two spectra have particularly high H$\alpha$ 
equivalent widths and show O I emission at 7773\AA \ and the higher orders of the Paschen series.

We determine the spectral types of our Hectospec and IMACS data with SPTCLASS, a semi-automatic 
spectral-typing program \citep{Hernandez04}. It uses empirical relations of spectral type and equivalent 
widths to classify stars. It has three schemes optimized for different mass ranges (K5 or later, late F to 
early K and F5 or earlier), which use different sets of lines. The user has to manually choose the best 
scheme for each star based on the prominent features in the spectrum and the consistency of several 
indicators. While SPTCLASS is insensitive to reddening and S/N of the spectra (as long as one can 
obtain a good flux estimate), it does not take into account the effect of the hot continuum emission 
produced by the accretion shocks. This continuum emission makes the photospheric absorption lines appear weaker. 
SPTCLASS generally assigns an earlier spectral type to veiled stars than their and 
therefore the SPTCLASS outputs should be considered as the earliest spectral type limits. 
Highly veiled stars (such as the ones shown in Figure~\ref{fig:spec_acc}) are not spectral-typed. 
Eight of our program stars are too veiled for spectral-typing.

\begin{figure}
	\begin{center}
		\includegraphics[scale=0.5]{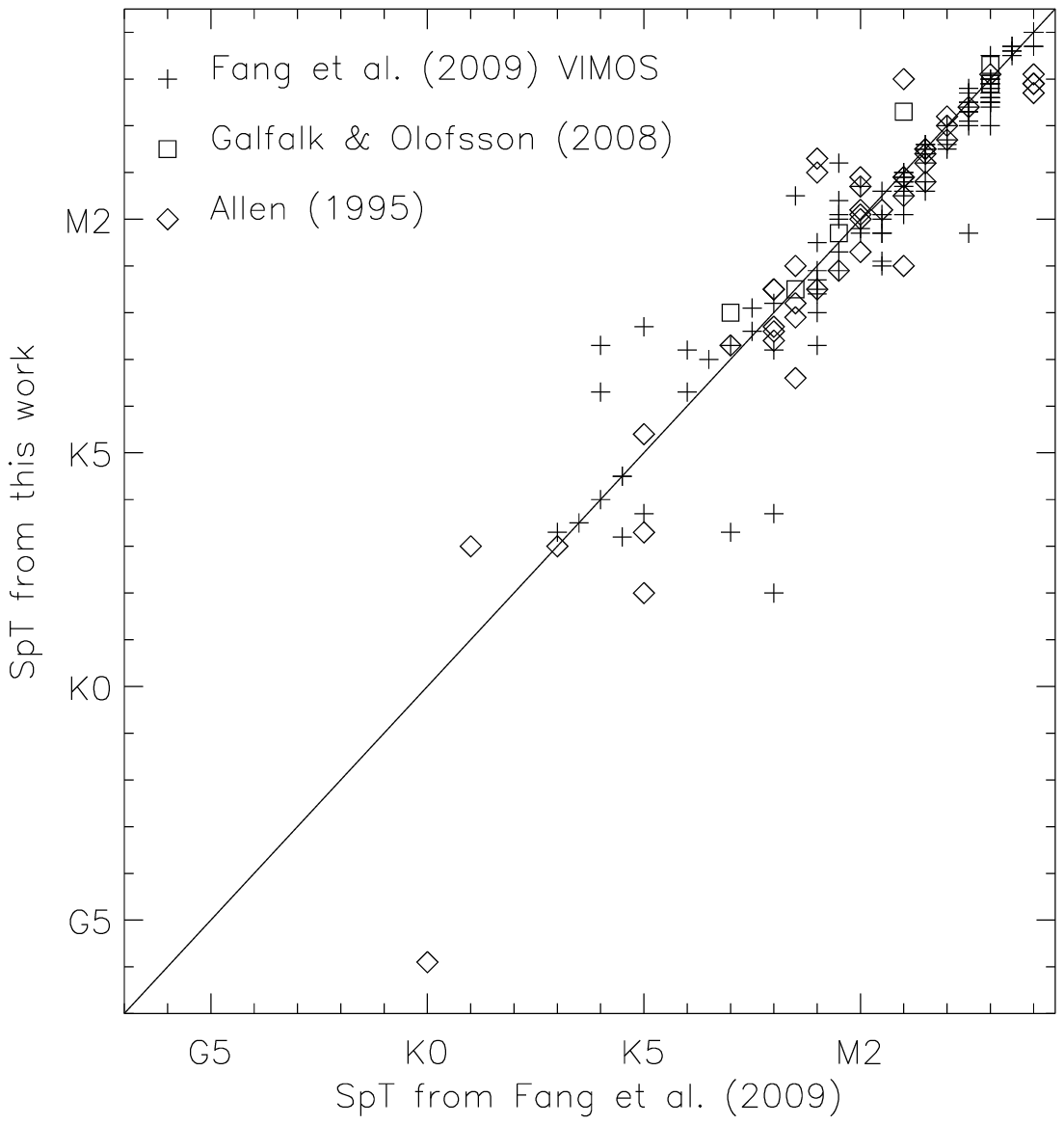}
	\end{center}
	\caption{Comparison of the spectral types obtained by \citet{Fang09} and this work for the 166 objects that are included in both samples. The catalog of Fang et al. (2009) comes from three sources: 1. their VLT/VIMOS observations (crosses) , 2. Galfalk\&Olofsson(2008) (squares)  and 3. Allen (1995) (diamonds).  The highly discrepant star near the bottom is heavily veiled (see text)}
	\label{fig:fang_compare}
\end{figure}

We also complement our results with data from Fang et al. (2009, hereafter F09),\nocite{Fang09}
which include spectral type estimates and H$\alpha$ and Li I equivalent widths for 266 stars. 
The majority of the spectral types come from their VLT/VIMOS 
observations and some are from \citet{Galfalk08} and \citet{Allen95}. 
We observed 206 of the stars in the F09 sample; 
Figure~\ref{fig:fang_compare} compares our spectral types with theirs.
Since the spectral types in F09 come from multiple sources, we use different
symbols to indicate the source.  
Most of the spectral type are consistent within 2 subclasses and there is no 
systematic trend in the differences; the only major 
large inconsistency is for a rapidly-accreting G type star, which has an
uncertain spectral type because of very high veiling.
For consistency we therefore use the spectral type from our observations for the objects in common,
and incorporate the data from F09 for the stars we did not observe.

\begin{figure}
	\begin{center}
		\includegraphics[scale=0.6]{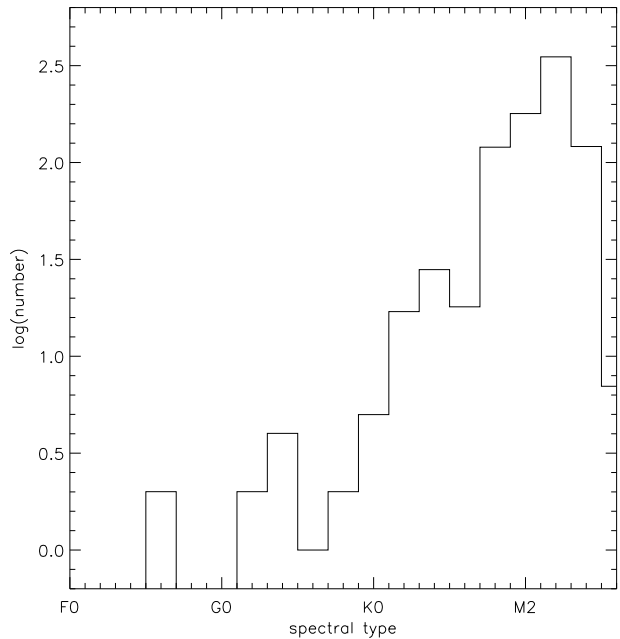}
	\end{center}
	\caption{Distribution of spectral types of all confirmed members with spectral type information from Table~\ref{tab:confirmed}. 
	Due to the magnitude criteria in our target selection, there are missing stars earlier than K in our sample. }
	\label{fig:spt}
\end{figure}

Figure~\ref{fig:spt} shows the distribution of spectral types of confirmed members. 
The distribution peaks around M4; we are systematically incomplete in the later-type 
population because of extinction (see \S 3.4). 
Due to the magnitude criteria in our target selection, 
there are very few stars earlier than K in our sample. We will address this
in a forthcoming paper, where we attempt to identify the intermediate-mass members.
 
 \begin{figure}
	\begin{center}
		\includegraphics[scale=0.55]{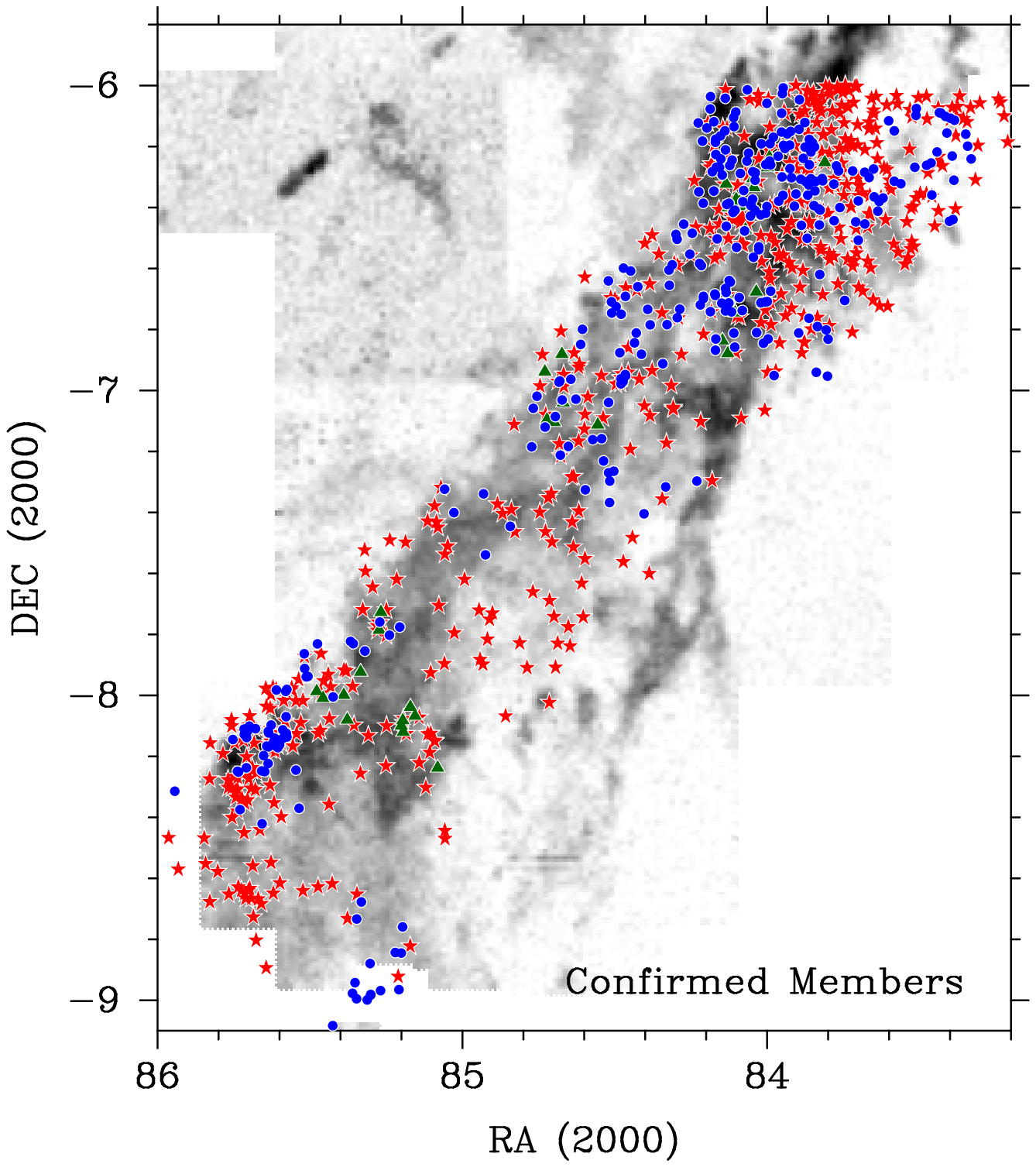}
		\includegraphics[scale=0.55]{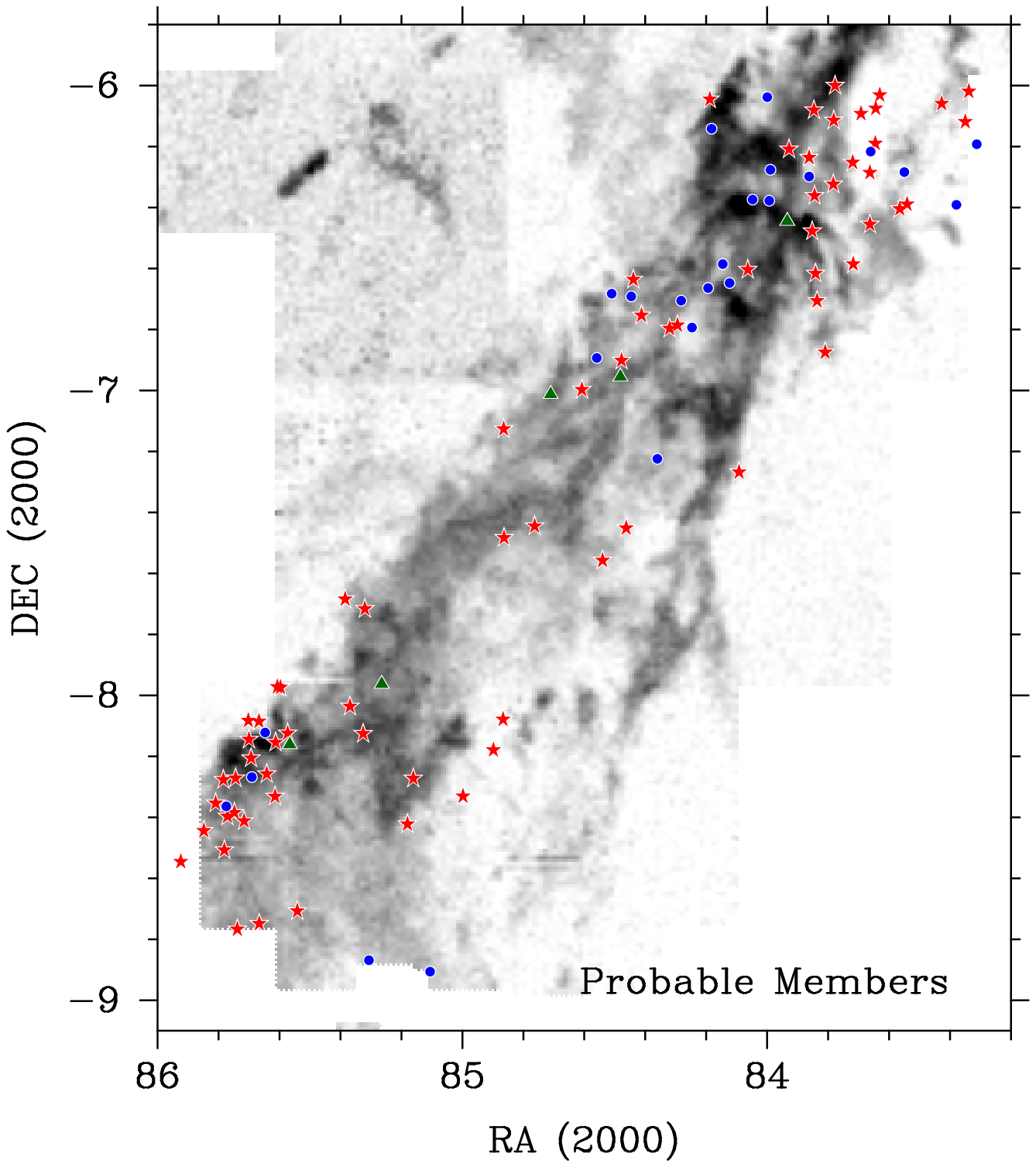}
	\end{center}
	\caption{Top: Positions of spectroscopically confirmed members in Table~\ref{tab:confirmed} overplotted on $^{13}$CO map from \citet{Bally87}. The symbols indicate
	the sources of the spectra. Red stars and blue circles represent objects observed with Hectospec and IMACS, respectively. Green triangles represent spectral types taken from F09.  Bottom: Positions of probable members in Table~\ref{tab:probables}. }
	\label{fig:loc}
\end{figure}

\begin{deluxetable*}{c|ccc|c|c}
\small
\tablecaption{Spectroscopic Members}
\tablecolumns{6}
\tablewidth{0pc}

\tablehead{\colhead{Source} & \colhead{Confirmed Members} & \colhead{IR excess} & \colhead{no IR excess} & \colhead{Probable} & \colhead{Total}}

\startdata

Hectospec & 517 & 235 & 282 & 69 & 586 \\
IMACS & 311 & 142 & 169 & 24 & 335 \\
Fang09\tablenotemark{a} & 36 & 29 & 7 & 5 & 41 \\
\hline
Total & 864 & 406 & 458 & 98 & 962 \\

\enddata	

\label{tab:stats}
 \tablenotetext{a} {Fang et al. (2009) cataloged 266 stars in L1641. The numbers here show only stars we did not observe. }
\end{deluxetable*}

We use the following criteria to identify young stars in our sample: 

(1) We include all IR excess stars using {\em Spitzer/IRAC} and 2MASS JHK colors 
(the IR-excess selection criteria can be found in 
\citealt{Gutermuth09}; the complete catalog will be published in Megeath \etal 2012).  
For all non-IR excess stars, we identify them as members if 
(2) Li absorption at $\lambda$6707\AA~is clearly detectable, and has an equivalent
width consistent with youth (the EW(Li) has to be greater than the Pleiades value for the
same spectral type from \citet{Briceno97}),
or (3) H$\alpha$ emission is clearly above the background nebulosity.

Criterion (2) is reliable in selecting pre-main sequence M and late K stars, but for early
K and G stars it can be more problematic, as Li depletion timescales can be quite long
(see, e.g., discussion in \citealt{Briceno97}).  Thus, to identify G and early K members,
we require that the equivalent widths of Li absorption lines to be greater than that observed
in Pleiades \citep{Briceno97} or the H$\alpha$ equivalent widths 
in emission or filled in above the expected absorption equivalent widths \citep{Stauffer97}.

Whether or not the observed H$\alpha$ emission - criterion (3) - is actually from the
star or simply nebular emission can be difficult to distinguish in some regions  and
especially for faint stars.  
We therefore compiled two lists of objects, one with only confirmed members of L1641 
and one that are potential members but have a higher contamination rate. 

To ensure a self-consistent sample, we extracted members from the F09 sample 
using the same criteria used in our own spectra; therefore not all the members listed in F09 
are included in our results here. In total, we adopted the F09 data for 41 stars that were not targeted in our
observations, five of them are considered ``probable members" due to the lack of Li absorption data. 

Table~\ref{tab:stats} lists the number of confirmed members (including IR-excess members and non-excess
members) and probable members from the three sources (Hectospec, IMACS, F09). 
Table~\ref{tab:confirmed} gives the positions, photometry, spectral types, etc. of the 864 confirmed 
members, confirmed either by their IR excess or unambiguous Li absorption. Out of 
the 864 confirmed members, 406 of them have IR excess, while 458 have no excess. Eight of the 
confirmed members are highly veiled by excess continuum emission, thought
to be produced by the accretion shock on the stellar photosphere \citep{Koenigl91}. 
Spectral typing is not possible for these objects.

Table~\ref{tab:probables} gives the positions, photometry, 
spectral types, etc. of the 98 probable members. 
The probable members are stars that show H$\alpha$ emission and are in the strip of V vs. V-I CMD 
(see bottom panel of Figure~\ref{fig:photometry}) that 
most of the confirmed YSOs reside in. They do not, however, have IR-excess or unambiguous Li absorption 
line (most likely due to low S/N of the spectra). We therefore catalog them as probable members. 
These probable members are excluded in our analysis and statistics unless otherwise noted. 

The top panel of Figure~\ref{fig:loc} shows the positions of spectroscopically confirmed members 
overplotted on the $^{13}$CO map from \citet{Bally87}. The distribution of the spectroscopic members in 
general follow the gas distribution. The bottom panel shows the positions of probable members. 
Probable members that are located near the group of confirmed members are more likely members 
whereas the ones far away from any confirmed members are likely non-members with nebular H$\alpha$ emission. 

\begin{figure}
	\begin{center}
		\includegraphics[scale=0.5]{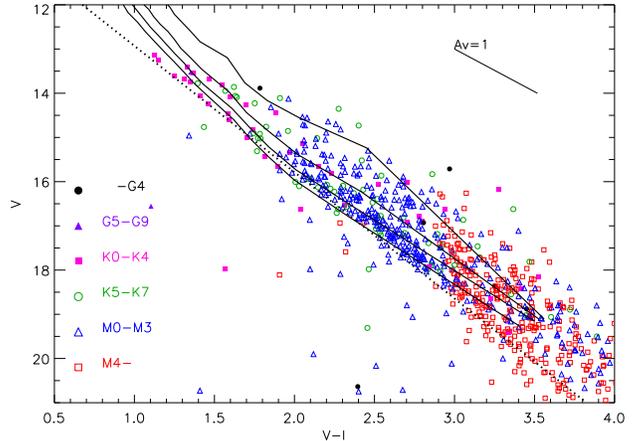}
	\end{center}
	\caption{Spectroscopically confirmed members on the CMD color-coded by their spectral types (not corrected for extinction). The solid lines are 1, 2, 3 and 4 Myr isochrones from \citet{Siess00} (from top to bottom) and the dotted line is the line used to define the YSO region. The minimum mass of the plotted isochrones is 0.1\Msun.}
	\label{fig:photo_members}
\end{figure}

Figure~\ref{fig:photo_members} shows the spectroscopically confirmed members on the CMD color-coded 
by their spectral types. It shows that we can spectral type objects down to M5-M6 as long as they are not highly extincted (
in general, $\rm{A_V}<2$). There are some objects well below the YSO regions and most of which are IR-excess objects. 
Most likely they are edge on disk objects that are seen in scattered light and therefore appear much bluer than
typical YSOs. This figure also shows that the magnitude limits of our optical photometry and spectroscopy 
are well-matched.

The solid lines in Figure~\ref{fig:photo_members}, from top to bottom, are the \citet{Siess00} isochrones
for 1, 2, 3 and 4 Myr. Because the isochrones are almost parallel to the reddening vector, we can use the
non-extinction corrected CMD to estimate the age of the population. Most of the population lies within the 1 Myr
and 4 Myr isochrone, and the distribution peaks around 3 - 4 Myrs. Using the Siess isochrones, 
\citet{DaRio10} found the age of the ONC to be $\sim$ 2 - 3 Myr.  The L1641 population, on average, 
thus appears to be slightly older than the ONC population, even though the age difference is comparable to the age uncertainties.  
This means that ultimately we should be able to make a direct comparison for IMFs using spectral
types without correction.

\tabletypesize{\scriptsize}
\begin{deluxetable*}{crrrrrrrrcrcc}

\tablewidth{0pt}
\tablecaption{Confirmed Members in L1641. This table is available only on-line as a machine-readable table. }
\tablehead{\colhead{ID} & \colhead{RA} & \colhead{Dec} & \colhead{V} & \colhead{I} & \colhead{J\tablenotemark{a}} & \colhead{H\tablenotemark{a}} & \colhead{K$_S$\tablenotemark{a}} & \colhead{Spectral Type} & \colhead{IR excess\tablenotemark{b}} &\colhead{EW(H$\alpha$)} &\colhead{EW(Li)} & \colhead{Date observed\tablenotemark{f}} \\
& \colhead{(J2000)} & \colhead{(J2000)} & \colhead{(mag) }& \colhead{(mag)} & \colhead{(mag)} & \colhead{(mag)} & \colhead{(mag)} & & & \colhead{(\AA)} & \colhead{(\AA)} & \colhead{}}

\startdata
  1&  83.21121&  -6.18593&    17.430&    14.793&    13.450&    12.740&    12.537&M3.0$\pm$0.6\tablenotemark{c}& N&     -4.30& 0.4&Oct 22, 2011\\
  2&  83.22298&  -6.10026&    17.670&    14.743&    13.146&    12.499&    12.259&M3.6$\pm$0.6\tablenotemark{c}& N&     -11.0& 0.4&Oct 22, 2011\\
  3&  83.23454&  -6.05524&    16.566&    14.356&    13.040&    12.405&    12.235&M1.5$\pm$0.7\tablenotemark{c}& N&     -6.00& 0.5&Oct 22, 2011\\
  4&  83.24131&  -6.04519&    16.819&    14.377&    12.975&    12.324&    12.117&M2.6$\pm$0.5\tablenotemark{c}& N&     -4.00& 0.3&Oct 22, 2011\\
  5&  83.30244&  -6.05771&    17.607&    14.925&    13.508&    12.879&    12.632&M3.0$\pm$0.6\tablenotemark{c}& N&     -4.20& 0.3&Oct 22, 2011\\
  6&  83.30896&  -6.19689&    17.006&    14.462&    13.006&    12.329&    12.074&M2.6$\pm$0.5\tablenotemark{c}& N&     -22.6& 0.3&Oct 22, 2011\\
  7&  83.31703&  -6.30980&    15.290&    13.320&    12.061&    11.383&    11.151&M0.5$\pm$0.5\tablenotemark{c}& N&     -3.60& 0.5&Oct 22, 2011\\
  8&  83.33062&  -6.24046&    17.858&    14.796&    13.158&    12.535&    12.274&M4.6$\pm$0.6\tablenotemark{d}& N&     -7.90& 0.3&Oct 11, 2011\\
  9&  83.33256&  -6.07309&    16.396&    14.081&    12.722&    12.013&    11.826&M2.2$\pm$0.6\tablenotemark{c}& Y&     -5.90& 0.4&Oct 22, 2011\\
 10&  83.34235&  -6.19836&    15.604&    13.608&    12.415&    11.724&    11.530&M0.6$\pm$0.5\tablenotemark{d}& N&     -3.20& 0.5&Oct 11, 2011\\

\enddata
\label{tab:confirmed}
 \tablenotetext{a} {\scriptsize J,H,K$_S$ photometry is from 2MASS}
 \tablenotetext{b} {\scriptsize Criteria for IR excess are defined in Gutermuth et al. (2009) and Megeath et al. (2012).}
 \tablenotetext{c} {\scriptsize Data from Hectospec spectra.}
 \tablenotetext{d} {\scriptsize Data from IMACS spectra.}
 \tablenotetext{e} {\scriptsize Data from \citet{Fang09}. No spectral-typing error estimate. }
 \tablenotetext{f}  {\scriptsize Local date at the beginning of the night.}

\end{deluxetable*}

\begin{deluxetable*}{crrrrrrrrcrcc}

\tabletypesize{\scriptsize}
\tablecaption{Probable Members in L1641. This table is available only on-line as a machine-readable table.}
\tablehead{\colhead{ID} & \colhead{RA} & \colhead{Dec} & \colhead{V} & \colhead{I} & \colhead{J\tablenotemark{a}} & \colhead{H\tablenotemark{a}} & \colhead{K$_S$\tablenotemark{a}} & \colhead{Spectral Type} & \colhead{IR excess\tablenotemark{b}} &\colhead{EW(H$\alpha$)} &\colhead{EW(Li)} & \colhead{Date observed\tablenotemark{f}} \\
& \colhead{(J2000)} & \colhead{(J2000)} & \colhead{(mag) }& \colhead{(mag)} & \colhead{(mag)} & \colhead{(mag)} & \colhead{(mag)} & & & \colhead{(\AA)} & \colhead{(\AA)} & \colhead{}}
\startdata

  1&  83.31244&  -6.19215&    18.897&    15.697&    14.001&    13.429&    13.142&M4.6$\pm$0.7\tablenotemark{d}& N&     -9.40& \nodata&Oct 11, 2011\\
  2&  83.33856&  -6.01960&    19.102&    15.608&    13.731&    13.171&    12.890&M4.6$\pm$1.0\tablenotemark{c}& N&     -11.6& \nodata&Oct 22, 2011\\
  3&  83.35003&  -6.11952&    16.900&    14.259&    12.708&    12.051&    11.804&M3.2$\pm$1.1\tablenotemark{c}& N&     -11.3& \nodata&Nov 20, 2006\\
  4&  83.37934&  -6.39085&    18.752&    15.434&    13.776&    13.197&    12.927&M5.0$\pm$0.7\tablenotemark{d}& N&     -12.7& \nodata&Oct 11, 2011\\
  5&  83.42722&  -6.05974&    19.915&    16.364&    14.475&    13.786&    13.501&M4.5$\pm$1.0\tablenotemark{c}& N&     -6.40& \nodata&Oct 22, 2011\\
  6&  83.54100&  -6.38943&    17.206&    14.619&    13.204&    12.576&    12.333&M3.1$\pm$0.5\tablenotemark{c}& N&     -6.90& \nodata&Oct 22, 2011\\
  7&  83.55022&  -6.28399&    18.653&    15.151&    13.345&    12.711&    12.427&M5.0$\pm$0.7\tablenotemark{d}& N&     -9.70& \nodata&Oct 11, 2011\\
  8&  83.56502&  -6.40473&    19.209&    15.796&    13.950&    13.362&    13.053&M4.0$\pm$1.0\tablenotemark{c}& N&     -5.70& \nodata&Oct 22, 2011\\
  9&  83.63090&  -6.03181&    19.917&    16.230&    14.216&    13.575&    13.287&M5.1$\pm$0.8\tablenotemark{c}& N&     -8.70& \nodata&Nov 20, 2006\\
 10&  83.64370&  -6.07540&    19.709&    16.125&    14.116&    13.593&    13.246&M5.0$\pm$2.0\tablenotemark{c}& N&     -8.70& \nodata&Oct 22, 2011\\

\enddata
\label{tab:probables}
 \tablenotetext{a} {\scriptsize J,H,K$_S$ photometry is from 2MASS}
 \tablenotetext{b} {\scriptsize Criteria for IR excess are defined in Gutermuth et al. (2009) and Megeath et al. (2012).}
 \tablenotetext{c} {\scriptsize Data from Hectospec spectra.}
 \tablenotetext{d} {\scriptsize Data from IMACS spectra.}
 \tablenotetext{e} {\scriptsize Data from \citet{Fang09}. No spectral-typing error estimate. }
 \tablenotetext{f}  {\scriptsize Local date at the beginning of the night.}
\end{deluxetable*}
\begin{deluxetable*}{rrrrrrrrrrrrrrrr}

\tabletypesize{\scriptsize}
\tablewidth{0pc}
\tablecaption{YSOs with strong emission lines. This table is available only on-line as a machine-readable table.}
\tablehead{\colhead{ID} & \colhead{RA} & \colhead{Dec} & \colhead{[O I]} & \colhead{[O I]} & \colhead{[O I]} & \colhead{[N II]} &\colhead{[N II]} &\colhead{He I} & \colhead{[S II]} &\colhead{[S II]} & \colhead{[O I]} &\colhead{[O I]} & \colhead{[Ca II]} & \colhead{[Ca II]} &\colhead{[Ca II]}\\
& \colhead{(J2000)} &\colhead{(J2000)} &\colhead{5577\AA} &\colhead{6300\AA} &\colhead{6363\AA} & \colhead{6548\AA} & \colhead{6583\AA}& \colhead{6678\AA}& \colhead{6716\AA}& \colhead{6731\AA}& \colhead{7773\AA}& \colhead{8446\AA}& \colhead{$8498\AA$}& \colhead{8542\AA}& \colhead{8662\AA}}

\startdata
  6&  83.30896&  -6.19689&     \nodata&     \nodata&     \nodata&     \nodata&     \nodata&     \nodata&     \nodata&     \nodata&     \nodata&     \nodata&   -0.56&   -0.27&   -0.12\\
  9&  83.33256&  -6.07309&     \nodata&    -2.0&   -0.37&     \nodata&     \nodata&     \nodata&     \nodata&     \nodata&     \nodata&     \nodata&     \nodata&     \nodata&     \nodata\\
 13&  83.35634&  -6.10948&     \nodata&    -5.5&    -1.4&     \nodata&   -0.40&    -1.4&     \nodata&     \nodata&   -0.70&    -1.9&    -6.8&    -6.5&    -5.8\\
 20&  83.38751&  -6.31044&    -4.7&    -81.&    -27.&     \nodata&    -7.7&    -1.1&    -5.6&    -8.7&    -1.0&    -5.6&    -15.&    -16.&    -13.\\
 29&  83.42479&  -6.26354&     \nodata&    -4.8&     \nodata&     \nodata&     \nodata&    -2.2&     \nodata&     \nodata&     \nodata&    -6.4&    -5.7&    -6.3&    -5.2\\
 34&  83.43911&  -6.07385&     \nodata&   -0.49&     \nodata&     \nodata&     \nodata&    -1.8&     \nodata&     \nodata&    -1.0&    -1.8&    -20.&    -22.&    -18.\\
 38&  83.45954&  -6.36361&     \nodata&    -2.7&   -0.70&     \nodata&    -1.5&     \nodata&   -0.16&   -0.47&     \nodata&     \nodata&     \nodata&     \nodata&     \nodata\\
 64&  83.57055&  -6.54707&     \nodata&    -4.7&    -1.2&     \nodata&     \nodata&     \nodata&     \nodata&     \nodata&     \nodata&     \nodata&     \nodata&     \nodata&     \nodata\\
 89&  83.65066&  -6.09297&     \nodata&   -0.75&   -0.13&     \nodata&     \nodata&    -1.4&     \nodata&     \nodata&   -0.59&    -1.4&    -2.0&    -2.1&    -2.0\\
 91&  83.65124&  -6.27344&     \nodata&    -1.3&     \nodata&     \nodata&     \nodata&    -2.2&     \nodata&     \nodata&     \nodata&     \nodata&     \nodata&     \nodata&     \nodata\\

\enddata
\label{tab:emission}
\end{deluxetable*}
\tabletypesize{\footnotesize}

\subsection{H$\alpha$~as a Membership Indicator and Accretion Diagnostics}\label{accretion}
 The optical spectra also provide us with diagnostics of accretion. Table~\ref{tab:confirmed} and~\ref
{tab:probables} list equivalent widths of H$\alpha$, which can be used as an indicator of accretion 
as the H$\alpha$ luminosity correlates with the accretion luminosity. \citet{Fang09} has an extensive discussion 
on accretion rates and disk properties in selected areas of L1641 using H$\alpha$, H$\beta$ and He I $\lambda$ 5876.

Table~\ref{tab:emission} lists other accretion indicator emission lines we found in the spectra. Note that if an 
emission line is not listed, it does not necessarily mean there is no emission line, but rather we cannot confidently tell 
that there is an emission line due to sky subtraction issues or limited spectral coverage in some IMACS data. 

 \begin{figure}
	\begin{center}
		\includegraphics[scale=0.7]{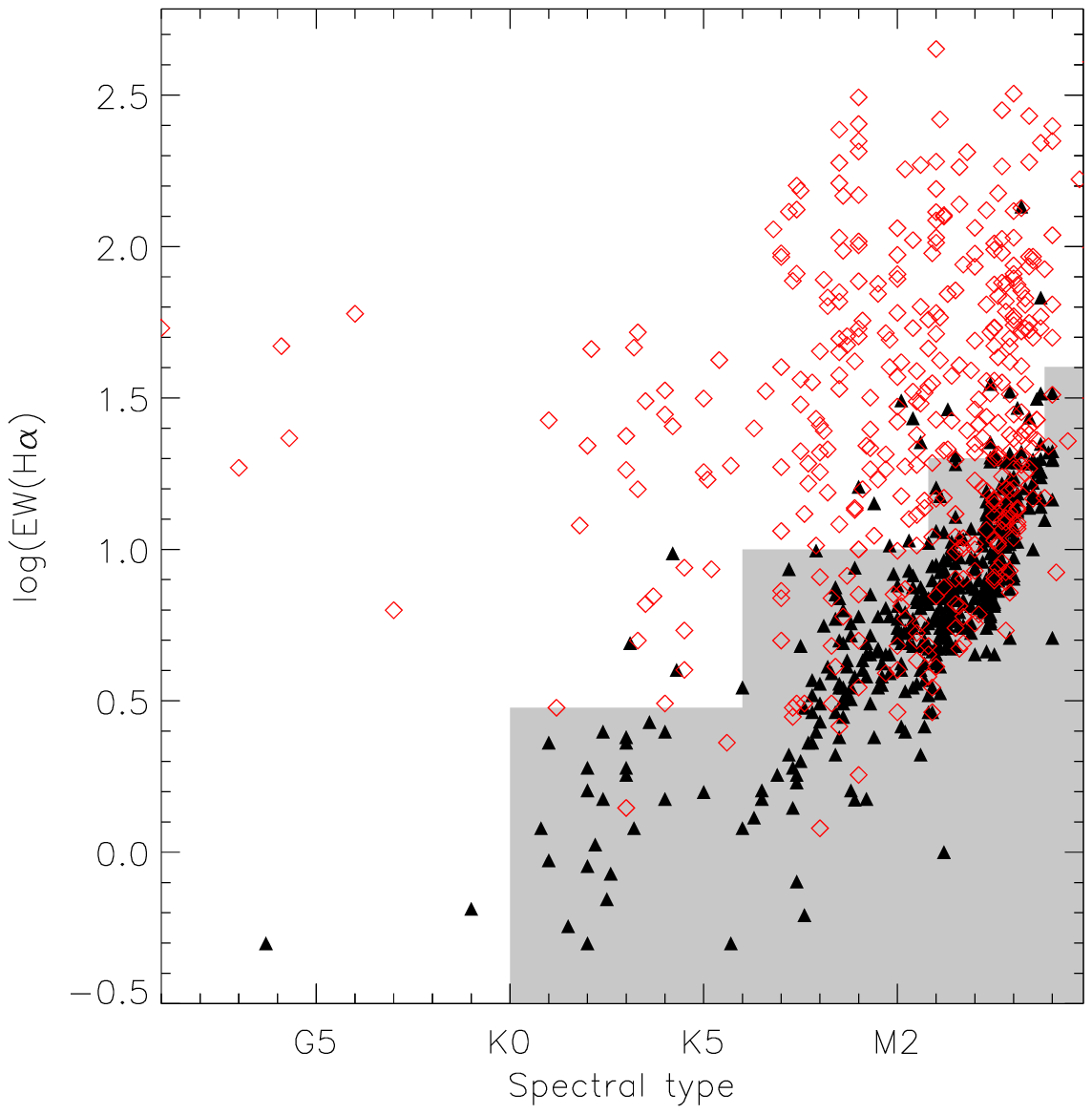}
	\end{center}
	\caption{H$\alpha$ equivalent widths (in log scale) of all confirmed members vs. spectral type. The red open diamonds are IR-excess members and black filled triangles are non IR-excess members. The shaded region shows where the WTTS objects reside using 
	CTTS/WTTS classification from White \& Basri (2003). Note that this classification only applies to stars later than K0.}
	\label{fig:Ha_Spt}
\end{figure}

  \begin{figure}
	\begin{center}
		\includegraphics[scale=0.7]{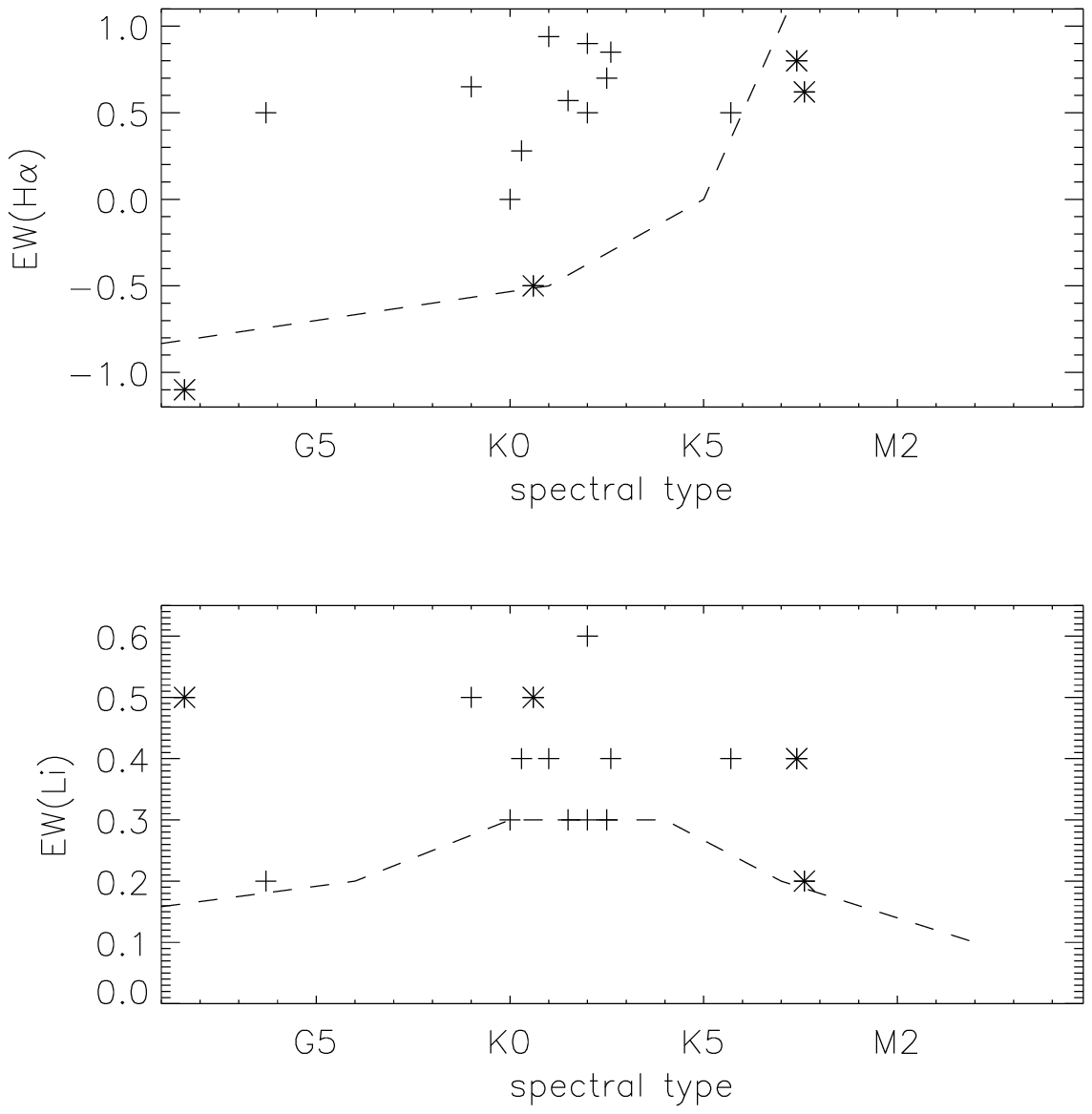}
	\end{center}
	\caption{Top:H$\alpha$ equivalent widths (in linear scale) of objects with H$\alpha$ equivalent widths less than 1\AA~in emission. The dashed line is the approximate relation for chromospheric activity in the Pleiades \citep{Stauffer97}. Note that here emission in H$\alpha$ is defined as positive equivalent width, consistent with Figure~\ref{fig:Ha_Spt}, but different from Table~\ref{tab:confirmed}. Bottom: Equivalent widths of Li absorption for the same objects. Absorption in Li is defined as positive equivalent width. The dashed line shows the approximate relation for the maximum Li absorption in the Pleiades for a given spectral type \citep{Briceno97}.}
	\label{fig:low_Ha}
\end{figure}

Figure~\ref{fig:Ha_Spt} shows H$\alpha$ equivalent widths of all confirmed members vs. spectral type. 
The H$\alpha$ equivalent widths can be used as an indicator of accretion as the H$\alpha$ luminosity
correlates with the accretion luminosity. In general, the H$\alpha$ equivalent widths have an increasing trend toward later spectral types and the 
IR-excess objects (red open diamonds) have higher H$\alpha$ \ equivalent widths than the non-excess objects (black
solid triangles). The shaded region shows the H$\alpha$ \ equivalent width criteria for CTTS and WTTS by \citet{WhiteBasri03}. Using this T Tauri criteria, 299 of the confirmed are CTTS and 565 are WTTS, which gives a CTTS/
WTTS ratio of 53\%. In the early K to G spectral type range, there are a few stars with very low EW(H$\alpha$) equivalent 
widths.

We now look at the stars with very low H$\alpha$ equivalent widths in more detail. 
The top panel of Figure~\ref{fig:low_Ha} shows the H$\alpha$ equivalent widths vs. spectral type for objects with H$\alpha$
equivalent widths less than 1\AA~in emission. Note that in this plot emission is positive and that the y axis is in linear scale. 
The dashed line show the H$\alpha$ equivalent widths measured in the Pleiades from chromospheric activity \citep{Stauffer97}.
We note that four of the members fall on or below the dashed line. The bottom panel of Figure~\ref{fig:low_Ha} shows the Li equivalent 
widths of the same stars, where the stars with particularly low H$\alpha$ are shown in asterisks. All the low-H$\alpha$ stars shown in the 
top panel show Li absorption above the Pleiades level \citep{Briceno97} and are listed as confirmed members. Although it is possible that some of these objects are from previous star-forming events in the Orion complex and still show H$\alpha$ and Li, the number of low H$\alpha$ objects is very small and has little effect on our overall sample.
\subsection{Extinction}\label{extinction}

We use the color excess E(V-I) to estimate the extinction toward a star. 
For stars earlier than M4, we use the main-sequence colors from \citet{KenyonHartmann95};
For stars M4 and later, we use the intrinsic colors of young disk population described by \citet{Leggett92}.
We then calculate the extinction assuming a standard extinction law with R$_V$ of 3.1 \citep{Cardelli89}. 
The E(V-I) method has certain limitations and uncertainties. For example, \citet{DaRio10} have shown 
that accretion onto a star makes it look bluer and therefore the extinction would be underestimated. 

\begin{figure}
	\begin{center}
		\includegraphics[scale=0.7]{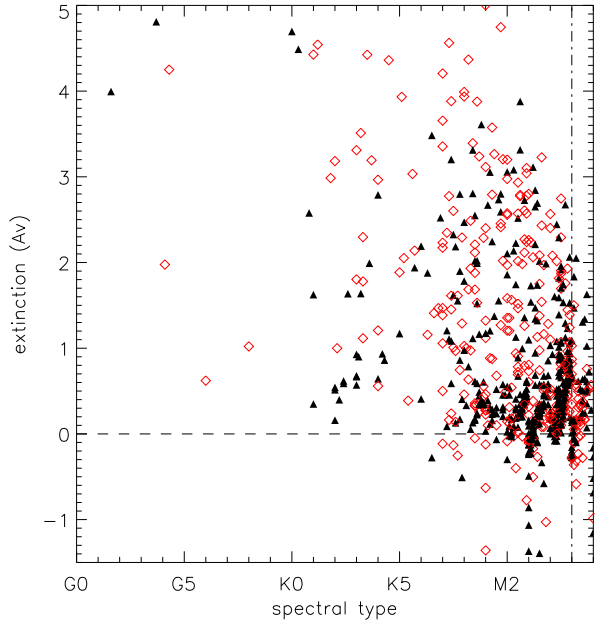}
	\end{center}
	\caption{Extinction vs. spectral types for IR excess objects (red open diamonds) and non-IR excess objects (black solid triangles) estimated from E(V-I), assuming the standard extinction law from \citet{Cardelli89} for R$_V$ of 3.1 and intrinsic colors from \citet{KenyonHartmann95} (for spectral types earlier than M4) and the young disk population in \citet{Leggett92} (for spectral types M4 and later).}
	\label{fig:extinction}
\end{figure}

Figure~\ref{fig:extinction} shows the distribution of extinction vs. spectral type . 80\% of our spectral-typed
members have \Av $\le$ 2. On average, the IR-excess objects (open red diamonds) appear to have higher extinction
than the non-excess objects, which is partially a selection effect because we are unable to identify non-excess members through 
Li absorption in high \Av~regions where the S/N is much lower for the same spectral type. Negative values of A$_V$s are non-physical and are due to errors in spectral-types, photometry and also uncertainties in their intrinsic colors. Since the spectral-typing errors and photometric errors are larger for the faintest objects, it makes sense to see more negative A$_V$s in the latest spectral types.

\begin{figure}
	\begin{center}
		\includegraphics[scale=0.5]{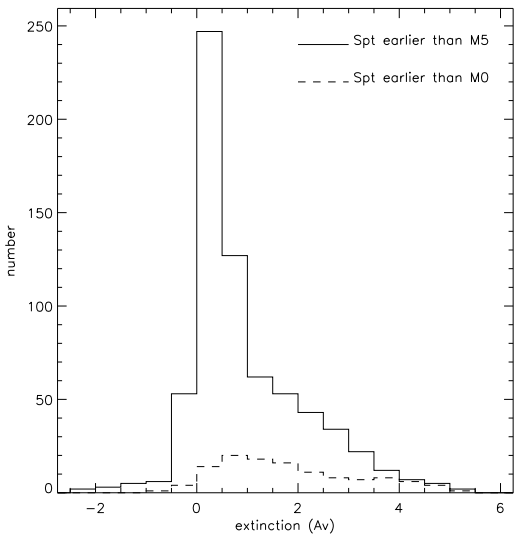}
	\end{center}
	\caption{Distribution of extinction toward members with spectral types earlier than M5 (solid line) and members with spectral types earlier than M0 (dashed line). This illustrates that our sample does not have a uniform cutoff in extinction. The extinction is estimated from E(V-I) assuming the standard extinction law from \citet{Cardelli89} for R$_V$ of 3.1 and intrinsic colors from \citet{KenyonHartmann95}. Stars later than M5 are not considered because of the large uncertainties in spectral-typing and photometry.}
	\label{fig:extinction_hist}
\end{figure}

Figure~\ref{fig:extinction_hist} shows the distribution of extinction of spectral-typed members earlier 
than M5 (solid line) and the subset of members earlier than M0 (dashed line). 
Stars later than M5 have large uncertainties in their spectral types and photometry and are 
excluded. The two distributions show that in extincted regions, our sample is biased 
toward more luminous objects because we did not apply an extinction cut in our sample selection.

\begin{figure}
	\begin{center}
		\includegraphics[scale=0.8]{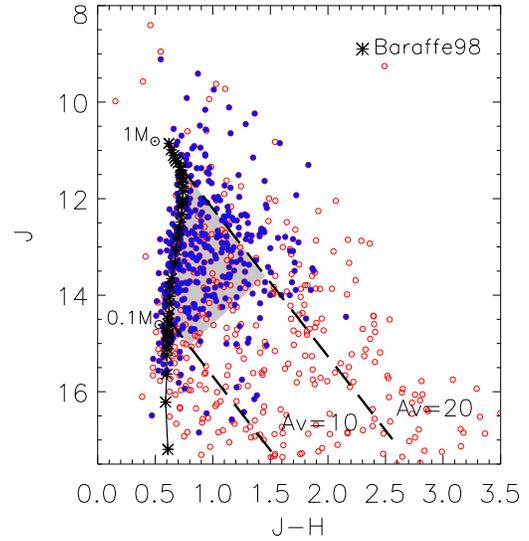}
		\includegraphics[scale=0.8]{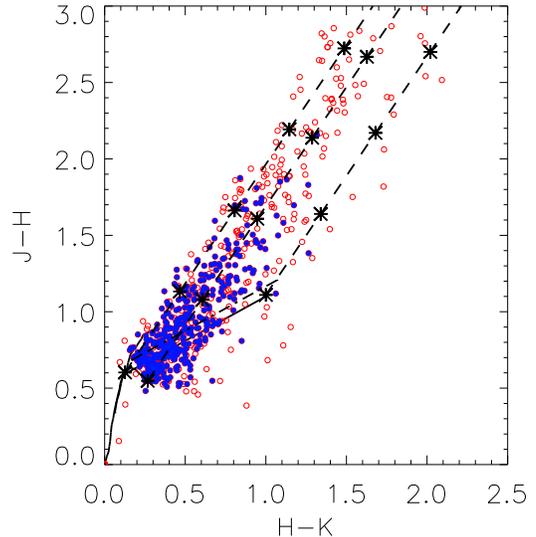}
	\end{center}
	\caption{Top: J vs. J-H color magnitude diagram of all IR-excess objects. The objects that we spectral-typed are shown in solid circles (blue) and objects without spectral types are shown in open circles (red). Overplotted are the 3 Myr isochrone from \citet{Baraffe98} and extinction vector from 1\Msun~(A$_V$ = 20) and 0.1\Msun~(A$_V$ = 10). In the shaded area, our optical spectroscopy sample is about 80\% complete. Bottom: J-H vs. H-K color-color diagram.  The two solid lines are the intrinsic colors of main sequence and giant branch \citep{BessellBrett88}. The dashed lines and the asterisks show the extinction in A$_V$ of 5 increments. In both plots, we use the standard extinction law from \citet{Cardelli89} for R$_V$ of 3.1.}
	\label{fig:JJH}
\end{figure}

\begin{figure}
       \begin{center}
		\includegraphics[scale=0.55]{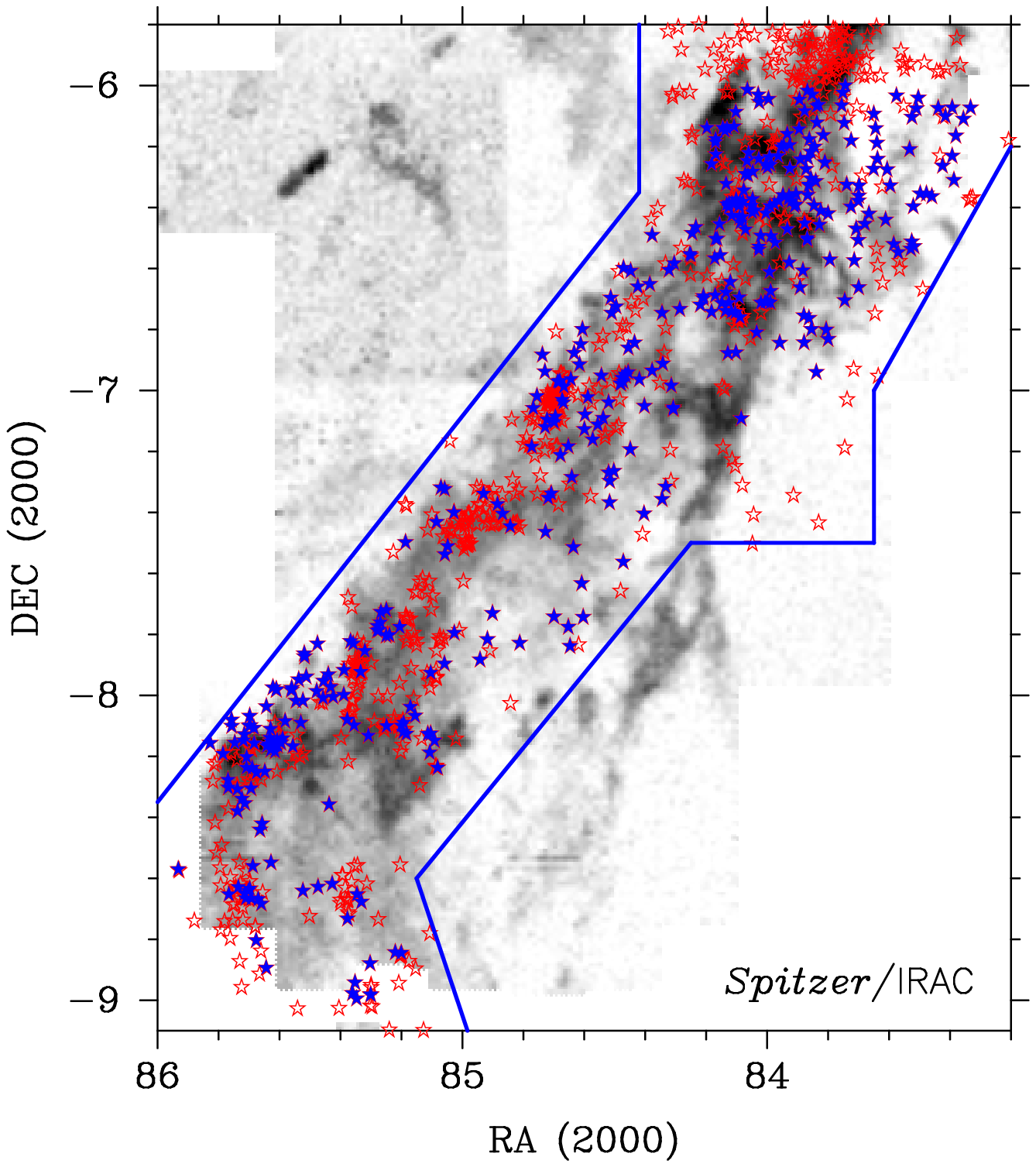}
	\end{center}
	\caption{Same as Figure~\ref{fig:spitzer_fields}, but the IR-excess stas with spectral types are plotted in blue solid symbols and 
IR excess stars without spectral types are plotted in red open symbols. Most of the missing spectral types are due to high extinction.}
	\label{fig:spitzer_spt}
\end{figure}

We now turn our attention to the IR-excess population and assess what percentage of this
population is highly extincted and not in our optical sample. 

We use the 2MASS colors of the IR excess disk objects to estimate the extinction and assess 
the completeness of our spectroscopic sample. 
The top panel of Figure~\ref{fig:JJH} shows J vs. J-H color magnitude diagram of all IR-excess objects. 
The objects that we spectral-typed are shown in solid circles (blue) and objects without spectral types 
are shown in open circles (red). 
Overplotted are the 3 Myr isochrone from \citet{Baraffe98} and extinction 
vector from 1\Msun~(A$_V$ = 20) and 0.1\Msun~(A$_V$ = 10). We use the isochrone from Baraffe et al. 
(1998) since they are appropriate for the mass range.
We assume an age of 3 Myr because our optical photometry (Figure~\ref{fig:photo_members}) shows that L1641
is approximately 3Mrys old (estimated using the \citet{Siess00} isochrones). 
and the 3 Myr \citet{Baraffe98} isochrone works the best to match up spectral type and J magnitudes. 
The bottom panel shows the J-H vs. H-K color-color diagram.  The two solid lines are the intrinsic colors of main 
sequence and giant branch \citep{BessellBrett88}. The dashed lines and the asterisks show the extinction 
in A$_V$ of 5 increments. In both plots, we use the standard extinction law from \citet{Cardelli89} for R$_V$ of 3.1

From the distribution of the spectral-typed objects, we found that our ability to determine spectral types is 
limited by both the mass (corresponding to the J magnitude) and extinction. For example we can identify and spectral
type 1\Msun~stars in dense regions up to A$_V$ = 8, but we can only identify and spectral type 0.1\Msun~stars 
up to A$_V$ = 2. There are 234 IR-excess sources in the shaded region of the top panel of Figure~\ref{fig:JJH} 
and more than 80\% of them (192) are spectral-typed. 
If we limit ourselves to objects with A$_V~<~2$ for 1 to 0.1\Msun, then our spectroscopic sample is about 90\% (90/110) complete. 
We expect similar completeness for non-excess objects. 
Some of the intermediate-mass and massive stars are missing as a result of 
brightness limit of our spectroscopic survey, which we will address in a forthcoming paper.
Of all the IR-excess
objects that lie between the 0.1\Msun~and 1\Msun~line, only 60\% of them have $A_V>2$, which generally
agrees with our previous finding that only half of the IR excess stars are visible in the optical with V$<$ 21. 

These results show that our spectral type distribution (Figure 10) is substantially incomplete at
the latest types.  This poses a problem for addressing the shape of the IMF at low masses, which
we will also address in the forthcoming paper.  However, from the point of view of determining
whether the {\em high-mass} end of the IMF is deficient relative to the low-mass end, 
missing low-mass stars from our sample only means that our results will be underestimates
of the significance of any depletion of high-mass stars.

Figure~\ref{fig:spitzer_spt} shows the distribution of IR-excess objects and whether we have their spectral 
type data. No spectra were obtained for objects above -6$^\circ$~since they are not considered part of L1641. South
of -6$^\circ$, we have spectral types for 50\% of the disk objects. Most of the missing spectral types 
are due to high extinction. In particular, the aggregates of stars between -7.6 - -6.8$^\circ~$ in declination are highly-extincted \citep{Gutermuth11} and therefore were too faint
for spectral typing. 

\begin{figure}
	\begin{center}
		\includegraphics[scale=0.6]{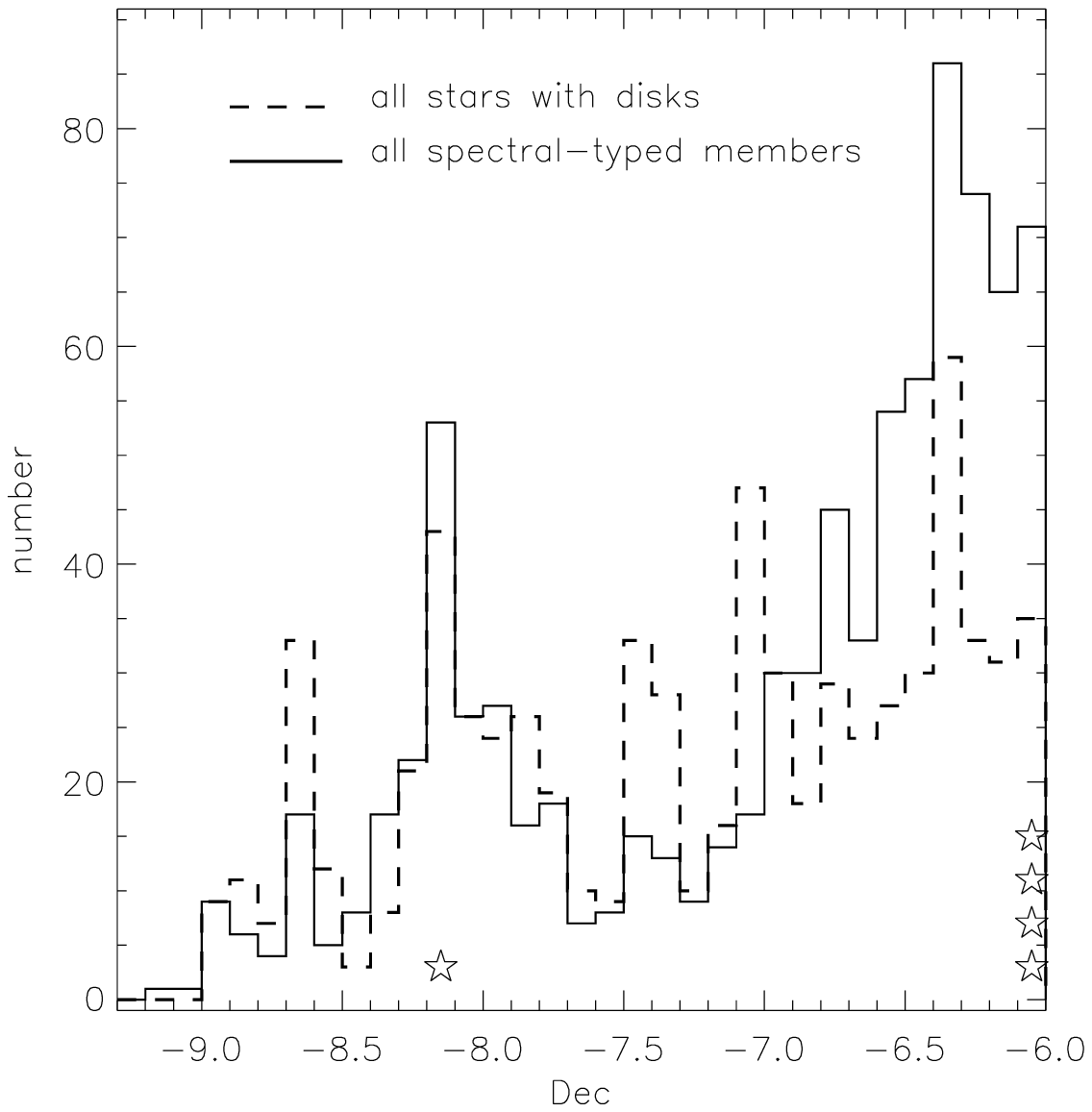}
		\includegraphics[scale=0.6]{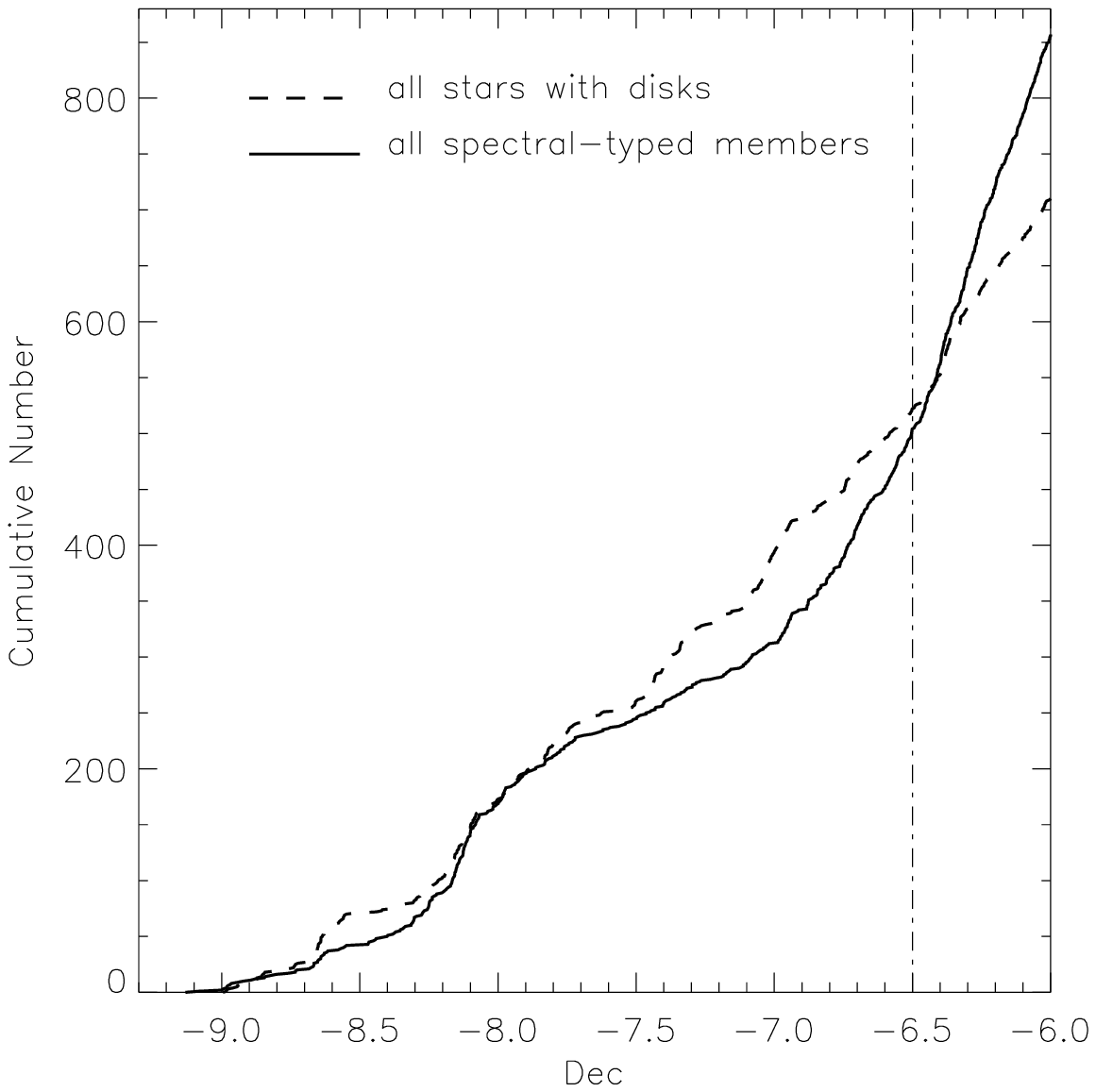}
	\end{center}
	\caption{Top: Distribution of spectroscopically confirmed, spectral-typed YSOs (solid line) and 
	all stars with disks (dashed line). The star symbols show the number of B0-B4 stars in each bin. The low number of spectral-typed stars around Dec = -7.5$^\circ$ - -6.8$^\circ$
	is a result of high extinction in this declination range. Bottom: Cumulative distribution of stars along increasing Declination. The horizontal dash-dot line represents a more strict cutoff of L1641 at Dec = -6.5$^\circ$ (see \S\ref{stats}). }
	\label{fig:Dec}
\end{figure}

Figure~\ref{fig:Dec} shows the distribution in Declination (top) and the cumulative distribution (bottom) of the optical
spectroscopically confirmed members (solid lines) and the distribution of all the IR-excess disk stars (dashed lines). 
In low extinction regions, we are able to classify most of the class II and class III's, and the number of spectroscopic 
members is larger than the IR excess sample; whereas in high extinction regions, there are very few optical members.
For example, the number of optical members is very small around Dec = -7.5$^\circ$ - -6.8$^\circ$ due to extinction. 

\section{Discussion}
\begin{deluxetable*}{c|cc|c}
\small
\tablecaption{Sample Definition}

\tablehead{\colhead{Definition}  & \colhead{IR excess disks} & \colhead{no IR excess} & \colhead{Total Number}}

\startdata

Spectroscopically confirmed members &  406\tablenotemark{d} & 458  & 864 \\

IR excess sample from Megeath \etal (2011) & 723 & & 723 \\
IR excess sample with spectral type & 406\tablenotemark{c} & & 406 \\
IR excess sample without spectral type & 317 &  & 317 \\
Class III's with moderate extinction \tablenotemark{a}& &$\gtrsim$458 & $\gtrsim$458 \\
All known members \tablenotemark{b}& 723 & 458 & 1181 \\

Total Number of Class III's\tablenotemark{c} &  & $\sim$900 & $\sim$900\\
\hline
Number of moderately extincted stars & 406 & $\gtrsim$458 & $\sim$ 860\\
Total number of stars & 723 & $\sim$900 & $\sim$1600 \\

\enddata	

\label{tab:sample}
 \tablenotetext{a} {Estimated number from the V vs. V-I CMD, minus the extrapolated foreground contamination. See \S~\ref{phot}.}
 \tablenotetext{b} {All known members include all spectroscopically confirmed members and the IR excess sample without spectral type.}
 \tablenotetext{c} {Assuming the class II/class III ratio is the same for the extincted population.}
 \tablenotetext{d} {This number include 8 objects that are classified as protostars, most of which are in fact flat-spectrum objects. }

\end{deluxetable*}

\subsection{Low-Mass Population}\label{sec:lowmass}
In \S~\ref{result_spec} we discussed our sample of optically confirmed members, including most of the moderately-extincted
class II's and some class III's that were observed serendipitously. In \S~\ref{extinction} we discussed the sample 
of highly extincted class II's that are without spectral types. In this section, we will estimate the number of class III objects 
expected in L1641 from the CMD and compare this number to the number of class III's that we have already observed.
Table~\ref{tab:sample} gives the definitions of the various populations and their numbers. (Note that numbers with $\sim$ signs are
estimates).

We noted that most of the IR-excess stars fall in a small region on the optical CMD (Figure~\ref{fig:photometry}) 
between the top two parallel lines (described in \S~\ref{phot}). 
We expect that the Class III objects have similar ages and distances and therefore 
fall in the same region of the CMD. However, not all the non-IR excess objects in this region are 
members. To estimate the fraction of the non-members in the non-IR excess sample, we use the strip 
one magnitude below the YSO region as a reference to estimate the number of foreground 
contaminants in the YSO region. Assuming no foreground extinction and uniform stellar density of 
foreground stars, the distribution of foreground stars in 
a given magnitude band is proportional to 10$^{0.6~m}$ (see \S~3.6 in \citealt{BinneyMerrifield}). 
Therefore, the number of foreground stars in the YSO region is predicted to be
$10^{-0.6}+10^{-1.2}$ times 
the number of foreground stars in the reference strip. 

We restrict our extrapolation to V-I range of 2 - 3.5 as this is the only 
portion where we have data for both the reference and the YSO region. 
We identified 514 stars in the reference strip between V-I of 2 - 3.5 
(only a very small fraction of which are members). 
The estimated contamination in the YSO region is therefore $514\times(10^{-0.6}+10^{-1.2})\approx$ 
162 stars. There are 496 non-IR excess objects in the YSO region between V-I of 2 to 3.5. We therefore
estimate that about 33\% of the non-excess stars in the YSO strip are non-members.

The entire YSO strip from V-I of  0.5 to 4 has 772 non-excess stars. If we take the same contamination
rate of 33\%, we estimate that $772\times(1-33\%)\sim520$ of these non-excess stars are members. 
This agrees roughly with the number of class III objects (458) that we confirmed spectroscopically and 
shows that we have identified the vast majority of the class III objects 
subject to our extinction limits. 

Approximately half (347 out of 723) of the IR-excess objects are identified in the optical photometry. 
If we further assume that the ratio of class II to class III's is constant even for the more extincted population, 
then we can estimate that there are approximately $458\times2\sim900$ class III objects. The total population 
of Class II + Class III stars in L1641 would then be roughly $\sim$ 1600. 

The above calculation, of course, is a very rough estimate, and depends upon our assumptions. First, we assumed 
that the the foreground stars are all main sequence stars and are uniformly distributed. We have also implicitly assumed that 
the class II to class III ratio is constant for all spectral types, while the observed disk frequency is a function of spectral 
type (see e.g. \citealt{Lada06}); however, this is probably a reasonable assumption for the majority of our members,
as the disk frequency seems to decrease most strongly for stars earlier than M.
The assumption that the disk frequency is the same for the extincted population can also be problematic. 
The {\em Spitzer/IRAC} survey has found that Class II objects in L1641 are more distributed spatially 
compared to the class I's \citep{AllenPPV}; similarly, it may be that the Class III objects are 
more distributed and have lower extinction systematically than the Class II stars. 
\citet{Megeath12} estimated a disk fraction of 70 - 80 \% in denser regions of L1641 by using off fields to subtract out the 
number of contaminating stars, higher than our the disk fraction we found for moderately extincted stars. Therefore, assuming
the same disk fraction could lead to an overestimate of the number of class III stars.

We can compare the number of low-mass stars in L1641 to the
results from surveys of the ONC. The ONC region has been surveyed multiple
times both in the IR and in the optical. The number of sources found highly depends on the area and depth of the 
specific surveys. 
Most of the IR surveys go much deeper than 2MASS, which makes comparing the number of sources more difficult. 
The optical survey of \citet{DaRio10} is probably the most comparable to ours since they have a 50\% V band completeness 
limit of 20.8. In an area of 34\arcmin~by 34\arcmin~they identified $\sim$ 1500 stars that are both present in the V and I bands.
\citet{Hillenbrand97} also has a similar survey area and found similar number of objects. 
Therefore, we conclude that L1641 contains comparable number of stars as the half degree field centered on the ONC to within a
factor of 2. This suggests that a direct comparison of the stellar
IMFs in the two regions can be statistically meaningful.

\begin{deluxetable*}{cccccccc}
\small
\tablecaption{List of early B-type Stars in L1641}

\tablehead{\colhead{\#}&\colhead{RA} & \colhead{Dec} & \colhead{Name} & \colhead{Spectral Type\tablenotemark{a}} & \colhead{V\tablenotemark{b}} & \colhead{B-V\tablenotemark{b}} & \colhead{In Spitzer Field?\tablenotemark{c}} \\
&\colhead{(J2000)} & \colhead{(J2000)} & & &\colhead{(mag)} &\colhead{(mag)} & \colhead{Y or N} }

\startdata
 1 &    83.7542    & -6.00928 &HD 36959 &B1.5V\tablenotemark{d}       &  	5.68	&-0.21 &Y\\
 2 &   83.7612    & -6.00203 &HD 36960 &B0.5V\tablenotemark{e}      &	4.79	&-0.25& Y\\
 3  &  83.8156    & -6.03275 &HD 37025 &B2/3V\tablenotemark{f}  	&	&&Y\\	
 4 &  84.1487    & -6.06475 &HD 37209 &B3IV\tablenotemark{g}   	&	5.72	&-0.22&Y\\
 5 &  84.6582    & -6.57397 &HD 37481 &B2V\tablenotemark{h}    &		5.96&	-0.22&N\\
 6 & 85.3434    & -6.93519 &HD 37889 &B2.5Vsn\tablenotemark{e} & 	7.65&	-0.1&N\\
 7& 85.5888    & -8.13339 &HD 38023 &B4V\tablenotemark{i} &    8.88&	0.33&Y\\

\enddata

\label{tab:early_B}
 \tablenotetext{a} {Most recent spectral type from \citet{Skiff10}.}
 \tablenotetext{b} {Optical photometry from \citet{WarrenHesser77}.}
 \tablenotetext{c} {The position on the sky is within the Spitzer field. We use the Spitzer field as a proxy of the main body of the cloud.}
 \tablenotetext{d} {The spectral type information comes from \citet{1990PASP..102..379W}}
 \tablenotetext{e} {The spectral type information comes from \citet{1977PASP...89..797A}}
 \tablenotetext{f} {The spectral type information comes from \citet{1983TrLen..38...62D}}
  \tablenotetext{g} {The spectral type information comes from \citet{2008ApJS..176..216A}}
  \tablenotetext{h} {The spectral type information comes from \citet{2009A&A...501..297Z}}
    \tablenotetext{i} {The spectral type information comes from \citet{Racine68}}

\end{deluxetable*}
\begin{figure}
	\begin{center}
		\includegraphics[scale=0.6]{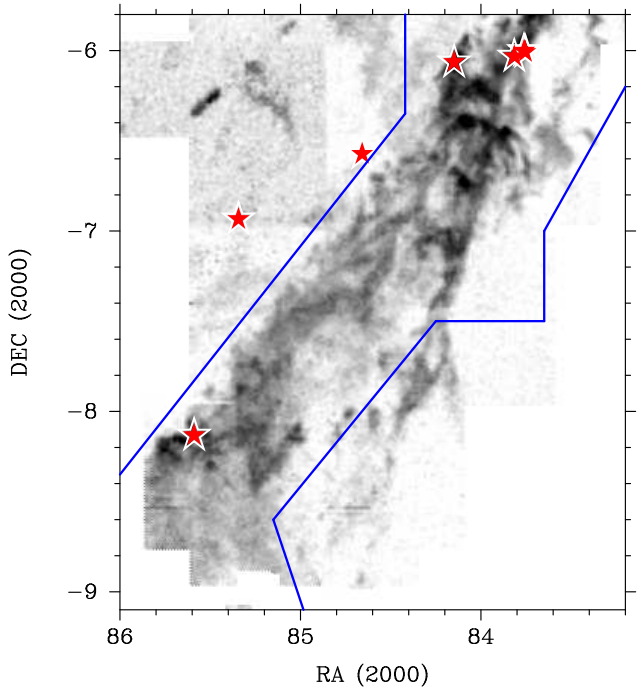}
	\end{center}
	\caption{Positions of known B0 - B4 stars near L1641 overlaid on the $^13$CO map \citep{Bally87}. The blue solid line outlines the \it{Spitzer/IRAC} fields. The spectral type information is taken from the spectral type compilation by \citet{Skiff10}. There are no O stars in this region.} 
	\label{fig:B_type}
\end{figure}

\begin{figure}
	\begin{center}
		\includegraphics[scale=0.5]{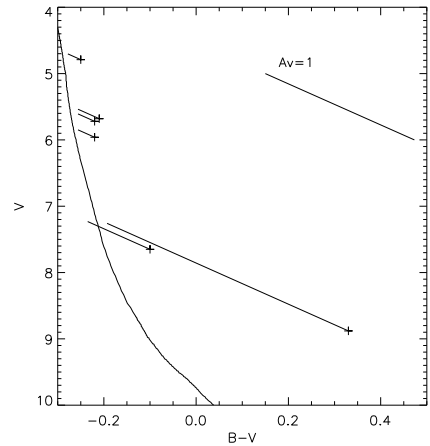}
	\end{center}
	\caption{ V. vs. B-V CMD of B type stars near L1641 (shown in Figure~\ref{fig:B_type}). The photometry is taken from \citet{WarrenHesser77}. The curved solid line is the ZAMS from Schaller et al. 1992 (Geneva tracks). Note that B and V photometry are not available for all B type stars. The solid line extending from each star shows its de-reddened position on the CMD, where the extinction is calculated from the E(B-V) color excess.}
	\label{fig:BV_early}
\end{figure}

\subsection{High-Mass Population}\label{sec:early_type}
With an idea of the number and spatial distribution of low-mass stars in L1641, we can now look into the 
high-mass end of the IMF. 
We search for known early type stars in L1641 in the catalog of spectral types compiled by \citet{Skiff10}.
There are no O stars in the area shown and the known early B type (B0 - B4) stars are shown in Figure~\ref{fig:B_type}, 
overlaid on the $^{13}$CO map\citep{Bally87}. Table~\ref{tab:early_B} lists the positions, spectral types and photometry of these stars.
In the case of spectral classifications from multiple sources, the most recent classification is used. There are a total of 
seven early B stars in this region. Four of the early B stars (1 - 4 in Table~\ref{tab:early_B})
are on the northern tip of the cloud and potentially denote the
outer regions of the ONC; two other stars (5 and 6) are not associated with the 
cloud if seen in $^{13}$CO emission but could be associated with the lower density gas probed by the near-IR (2MASS)
extinction map. Since we do not have information on the low-mass population in their surroundings and they are not clearly 
associated with L1641, we will not consider these two stars in the IMF analysis. 
The other early B star (7) is the B4 star near (RA,Dec) = (5:42:21.3, -08:08:00) \citep{Racine68}. 
Figure~\ref{fig:BV_early} shows the V vs. B-V CMD of these early B stars with photometry from Warren and Hesser (1977) \nocite
{WarrenHesser77} with the ZAMS from \citealt{Schaller92} (Geneva tracks) overplotted. The dereddened position of each 
star is also shown, where the extinction is calculated from the E(B-V) color excess. 
All the known early B stars are consistent with the ZAMS at the distance of 420 pc, which means they are likely 
associated with Orion A. 

From Figure~\ref{fig:BV_early}, we found that all of the known early B stars have small extinctions of A$_V <$ 2, which is a 
result of the depth of the \citet{WarrenHesser77} photometry. It is therefore possible that there are embedded O or early B 
stars that have not been found. However, based on the existing {\em Spitzer} survey \citep{Megeath12}, it is unlikely that 
we miss an early B star. Evolved O or early-B stars would have blown away materials nearby in a manner similar to the 
B4 star near (RA, Dec) = (5:42:21.3, -08:08:00) and therefore will not have high extinction. The reflection nebula near the 
B4 star is the only one seen in our field, which supports that there are no evolved early-B stars. (Note that there is another 
reflection nebula, powered by a potentially massive star further south, outside of the fields of our optical studies.) On the 
other hand, if there were unevolved, highly-extincted early-B stars, they would probably heat up the dust nearby 
and be identified as bright IR-excess sources by the {\em Spitzer} survey.
We first examine possible high-mass disk objects in the J vs. J-H CMD (top panel of Figure 17). The 
brightest star in the J band is on the upper left corner and has a spectral type of A1. A few other stars appear to have similar 
mass and are more extincted, but they all have similar masses according to the diagram. The only exception is the FU 
Ori star V883 Ori (with J = 9.25 and J-H = 2.49), which is dominated by accretion luminosity. We then examine the protostars
that have large bolometric luminosities. The most luminous protostar in L1641 is located at (RA, Dec) = (5:40:27.4,-7:27:30)
in one of the high-extinction clumps. It has L $\sim$ 490 \Lsun \citep{Kryukova12}, and according to the Siess (2000) isochrones, 
has a mass less than 7 \Msun. 

\subsection{Comparing the High-Mass IMF of L1641 to Analytical IMFs}\label{stats}

In \S~\ref{sec:early_type}, we discussed the high mass population in L1641 from a literature search. In our following
analysis, we assume that there are no other early B stars in the region. All but one of the early B stars 
are in the northern region contiguous with the ONC region. This raises the question of how we should 
define the region in which we construct the IMF. We need to adopt a more conservative definition and 
limit our sample to stars from a lower declination. This also means we have to compromise
for a smaller sample size. Figure~\ref{fig:Dec} shows that a large number of the confirmed members lie in the 
northern end of the cloud and that our sample size will be considerably smaller if we consider only the members
further south. We need choose carefully a cutoff for our L1641 sample. 

Here we discuss whether this lack of high mass stars is statistically significant with the following two approaches: 
1. considering the entire L1641 region south of -6$^\circ$ \ and 2. considering only the stars south of -6.5$^\circ$. 
For comparison, we also consider the region between -6$^\circ$~and -6.5$^\circ$, which is a relatively higher density
region in L1641 and contains 4 early B stars. Given the sample size n and the mass of the most massive star in the region \Mmax, 
we can then calculate how unlikely that this sample is drawn from analytic IMF models. Here we only consider the \citet{Chabrier05} 
and \citet{Kroupa01} IMFs. The \citet{Chabrier03} IMF gives results that are in between the two IMFs discussed here. 

First we normalize the IMFs between 0.1 and 120\Msun. The \citet{Chabrier05} IMF can be written as: 
\begin{equation}
\xi(\log m) =
\begin{cases} 
	=  A\;\exp\{-\frac{(\log m - \log 0.25)^2}{2\times (0.55)^2}\}, & 0.1M_\odot < m < 1 M_\odot, \\
	=  0.549A\;m^{-1.35},  &m \ge 1 M_\odot,
\end{cases}
\label{eq:IMF2}
\end{equation}
where A $\sim$ 0.959 normalizes the total number to unity.  

The \citet{Kroupa01} IMF can be written as: 
\begin{equation}
\xi(m) 
\begin{cases} 
	=  2A\;m^{-1.3}, & 0.1M_\odot < m < 0.5 M_\odot, \\
	=  A\;m^{-2.3},  &m \ge 0.5 M_\odot,
\end{cases}
\label{eq:IMF}
\end{equation}
where the normalization constant A $\sim$ 0.1431. 

The cumulative distribution function is
\begin{eqnarray}
\Sigma (M) & =  &\int_{\log (m = 0.1\rm{M}_\odot)}^{\log M} \xi(\log m)~d \log m \nonumber \\
		  & =  &\int_{m = 0.1\rm{M}_\odot}^{M} \xi(m)~dm
\end{eqnarray}

The probability that a randomly sampled star from this IMF is more massive than \Mmax~is 
$p = 1-\Sigma($\Mmax). In a population of n stars, the expected number of stars more
massive than \Mmax~is $E(>$\Mmax$)=np$. The probability that the most massive star is 
below \Mmax~is $P = (1-p)^n$. 

Table~\ref{tab:probability} lists the sample size n, the most massive star in the region \Mmax and 
$p = 1-\Sigma($\Mmax), $E(>$\Mmax$)=np$ for both \citet{Chabrier05} and \citet{Kroupa01} IMFs.
Note that here we consider two sample sizes: spectral-typed members and total known members, 
which is the sum of spectral-typed members and the IR-excess members that 
are too extincted for optical spectroscopy. Generally, the \citet{Chabrier05} IMF gives a higher number
of high-mass stars and therefore it is less likely to find no high-mass stars. 

If we consider the entire L1641 region south of -6$^\circ$, the most massive star is a B0.5V star (corresponding
to \Mmax$\sim$ 16\Msun). We expect to see 2.4 to 4.4 stars more massive than the B0.5V star in such 
a population. If the populations are randomly drawn, the probability of not finding any star more massive than 
16\Msun is quite low (0.012 to 0.091). Therefore, this population has fewer high-mass stars compared to 
the analytic IMF models, but the result is not very statistically significant. 

If we consider only the region between -6$^\circ$~and -6.5$^\circ$, the most massive star in this region (B0.5V)
is compatible with random sampling of both the analytical IMFs. 

By limiting our sample to stars south of -6.5$^\circ$ and on the cloud (i.e. within the {\it Spitzer} survey area), 
we can ensure that our sample is not confused with stars from the ONC. There are no stars earlier than B4 
(corresponding to \Mmax$\sim$ 7\Msun) in this sample. We expect to see 4.3 to 9.6 stars more massive than 
the B4V star in such a population. If the populations are randomly drawn, the probability of not finding any 
star more massive than 7\Msun is very low ($1.4\times 10^{-2}$ to $6.4\times10^{-5}$). If we consider only the 
500 members with optical spectral-types, the population is incompatible with the analytic IMFs to a 2 - 3 $\sigma$
level; if we further include the 262 IR-excess members without optical spectral-type, the population is
incompatible with the analytic IMF to a 3 - 4 $\sigma$ level.

Even though we currently cannot constrain the total number of highly extincted stars, the total population will only be larger if we 
consider highly-extincted class III stars. Assuming a complete sample of early B stars and the lower limit of low-mass 
stars, the probabilities we calculated above are upper limits. Therefore the existence of highly-extincted low mass stars will 
only strengthen and not weaken our conclusions. On the other hand, our conclusion relies very strongly on the assumption that
there are no highly-extinced high mass members, especially stars earlier than B4 in the region south of Dec = -6.5$^\circ$.
We also want to point out that the expected number of high-mass stars varies greatly for different IMF models due to their 
discrepancies in the low-mass end. The low-mass IMF is less well-studied and the discrepancies in the models can be viewed as 
an indication of the uncertainties. 

Since we do not yet have a complete analysis of the high-mass and intermediate-mass stars in L1641, we choose not to 
present an HR diagram of the low-mass sample alone. We will present the high-to-intermediate mass 
sample in a forthcoming paper, in which we will be able to construct a full IMF (down to M3) of L1641 and compare it to that of the ONC. 

\begin{deluxetable*}{c|c|c|c|c|c}
\small
\tablecaption{Samples by Region and IMF Probabilities}

\tablehead{\colhead{Region}  & \colhead{n (Sample Size)} & \colhead{M$_{\rm{max}}$\tablenotemark{a}} & \colhead{$p = 1 - \Sigma(\rm{M}_{\rm{max}})$\tablenotemark{b}} & 
 \colhead{$E(>\rm{M}_{\rm{max}})$\tablenotemark{c}}  & \colhead{P(no stars greater than M$_{\rm{max}}$)\tablenotemark{d}}}
\startdata

L1641						& Spectral-typed members 864					& B0.5V 	& Chabrier05 $3.74\times10^{-3}$	& 3.2		& 0.039	\\ [1ex]
Dec $<$ -6$^\circ$				& 										& (16\Msun)				& Kroupa01 $2.77\times10^{-3}$	& 2.4		& 0.091	\\ [1ex]
\cline{2-6}
							& Total Known Members\tablenotemark{e}1181	& B0.5V 	& Chabrier05  $3.74\times10^{-3}$	& 4.4		& 0.012	\\ [1ex]
							& 										& (16\Msun)				& Kroupa01  $2.77\times10^{-3}$	& 3.3		& 0.038	\\ [1ex]
\hline
							& Spectral-typed members 364					& B0.5V 	& Chabrier05  $3.74\times10^{-3}$	& 1.4		& 0.26	\\ [1ex]
Dec =  -6$^\circ$~to -6.5$^\circ$	&										&(16\Msun)				& Kroupa01 $2.77\times10^{-3}$	& 1.0		& 0.36	\\ [1ex]
\cline{2-6}
							& Total Known Members\tablenotemark{e} 419	& B0.5V 	& Chabrier05 $3.74\times10^{-3}$	& 1.6		& 0.20	\\ [1ex]
							&										& (16\Msun)				& Kroupa01 $2.77\times10^{-3}$	& 1.2		& 0.31	\\ [1ex]
\hline
							& Spectral-typed members 500					& B4V 	& Chabrier05 $1.26\times10^{-2}$	& 6.3		& $1.8\times10^{-3}$ \\ [1ex]
Dec $<$ -6.5$^\circ$				&										& (7\Msun)				& Kroupa01 $8.55\times10^{-3}$	& 4.3		& $1.4\times10^{-2}$ \\ [1ex]
\cline{2-6}
							& Total Known Members\tablenotemark{e} 762	& B4V 	& Chabrier05 $1.26\times10^{-2}$	& 9.6		& $6.4\times 10^{-5}$ \\ [1ex]
							&										& (7\Msun)				& Kroupa01 $8.55\times10^{-3}$	& 6.5		& $1.4\times 10^{-3}$
\enddata	
\tablenotetext{a} {Most massive star observed in this region.}
\tablenotetext{b} {The probability that a random star drawn from the IMF is more massive than M$_{max}$.}
\tablenotetext{c} {Expected number of stars more massive than M$_{max}$. $E(>\rm{M}_{\rm{max}}) = np$}
\tablenotetext{d} {Drawing n stars from the given IMF, the probability of getting no stars more massive than M$_{\rm{max}}$. P = (1-p)$^n$.}
\tablenotetext{e} {Total known members include spectral-typed members and IR excess members without spectral type.}
\label{tab:probability}
\end{deluxetable*}

\section{Conclusions}

We conducted an optical photometric and spectroscopic survey of L1641 to test whether IMF varies with the
environmental densities. 
Our spectroscopic sample consists of IR-excess objects selected from the {\em Spitzer/IRAC} 
survey and optical photometry and non-excess objects selected from optical photometry. We have spectral-typed 406
IR-excess members. We have also identified and spectral-typed 458 non-excess YSOs and 98 probable members 
through Li absorption and H$\alpha$ emission. Our sample is limited by extinction and stellar mass in a complicated 
way, but overall we are able to spectral-type about 90\% of class II and class III YSOs between 0.1 and 1 \Msun \ with A$_V\lesssim$ 2. 
The number of spectroscopically confirmed class III objects is also consistent with the number of class III objects 
estimated from extrapolation of the V vs. V-I color magnitude diagram. 

Assuming the class III/class II ratio is the same for the more extincted population, we estimate the total number of
class II and class III's in L1641 to be around 1600, including 723 class II objects from the {\em Spitzer} survey and $\sim$900 class III's.

The total number of stars in L1641 is comparable within a factor of two to the number of stars used to construct the IMF in the ONC, 
even though the number depends on how our L1641 sample is defined. The optical photometry shows that the ONC and L1641 are of similar age, with L1641 
being slightly older by $\sim$1 Myrs, indicating that any late-O to early-B star should still be on the main sequence. 
Assuming that we know all the early B stars in L1641, we compare the high-mass IMF to the standard analytical models. Compared to the standard models of the IMF, L1641 is 
deficient in O and early B stars to a 3-4$\sigma$ significance level. 
We will make further improvements to our sample to better constrain the extincted population and compile a complete sample of 
B stars in a forthcoming paper, and the we will be able to make a direct comparison with the ONC.

\acknowledgments
We thank the referee for helpful suggestions to improve the manuscript.  W.H. and L.H. acknowledge the support of NSF grant AST-0807305 and the Origins 
grant NNX08AI39G. This paper uses data obtained at the MMT Observatory, a joint facility of the Smithsonian Institution and the University of Arizona and
the 6.5 meter Magellan Telescopes located at Las Campanas Observatory, Chile.  This paper uses data products produced by the OIR Telescope Data 
Center, supported by the Smithsonian Astrophysical Observatory. 

{\it Facilities:} \facility{MMT (Hectospec)}, \facility{Magellan:Baade (IMACS)}, \facility{Hiltner (OSMOS)}, \facility
{Spitzer (IRAC)},\facility{FLWO:2MASS},

\bibliography{./Bib}

\end{document}